\documentstyle[12pt,epsfig,graphicx,color,amsmath, amsthm]{article} 
\makeatletter \def\@cite#1#2{{[{#1}]\if@tempswa\typeout {IJCGA
warning: optional citation argument ignored: `#2'} \fi}}

\newcount\@tempcntc
\def\@citex[#1]#2{\if@filesw\immediate\write\@auxout{\string\citation{#2}}\fi
  \@tempcnta\z@\@tempcntb\m@ne\def\@citea{}\@cite{\@for\@citeb:=#2\do
    {\@ifundefined
       {b@\@citeb}{\@citeo\@tempcntb\m@ne\@citea\def\@citea{,}{\bf ?}\@warning
       {Citation `\@citeb' on page \thepage \space undefined}}%
    {\setbox\z@\hbox{\global\@tempcntc0\csname b@\@citeb\endcsname\relax}%
     \ifnum\@tempcntc=\z@ \@citeo\@tempcntb\m@ne
       \@citea\def\@citea{,}\hbox{\csname b@\@citeb\endcsname}%
     \else
      \advance\@tempcntb\@ne
      \ifnum\@tempcntb=\@tempcntc
      \else\advance\@tempcntb\m@ne\@citeo
      \@tempcnta\@tempcntc\@tempcntb\@tempcntc\fi\fi}}\@citeo}{#1}}
\def\@citeo{\ifnum\@tempcnta>\@tempcntb\else\@citea\def\@citea{,}%
  \ifnum\@tempcnta=\@tempcntb\the\@tempcnta\else
   {\advance\@tempcnta\@ne\ifnum\@tempcnta=\@tempcntb \else 
\def\@citea{--}\fi
    \advance\@tempcnta\m@ne\the\@tempcnta\@citea\the\@tempcntb}\fi\fi}
\makeatother

\def\boxit#1{\leavevmode\thinspace\hbox{\vrule\vtop{\vbox{\hrule%
        \vskip3pt\kern1pt\hbox{\vphantom{\bf/}\thinspace\thinspace%
        {\bf#1}\thinspace\thinspace}}\kern1pt\vskip3pt\hrule}\vrule}%
        \thinspace}
\def\Boxit#1{\noindent\vbox{\hrule\hbox{\vrule\kern3pt\vbox{
\advance\hsize-7pt\vskip-\parskip\kern3pt\bf#1 \hbox{\vrule height0pt
depth\dp\strutbox width0pt} \kern3pt}\kern3pt\vrule}\hrule}}

\newcommand{\gsim}{\lower.7ex\hbox{$\;\stackrel{\textstyle>}{\sim}\;$}}
\newcommand{\lsim}{\lower.7ex\hbox{$\;\stackrel{\textstyle<}{\sim}\;$}}

\allowdisplaybreaks[1]

\def\bad{\begin{aligned}[t]}
\def\ead{\end{aligned}}
\def\eqnum{\stepcounter{equation} \tag{\theequation}}

\def\ifmath#1{\relax\ifmmode #1\else $#1$\fi}
\def\mgut{\ifmath{M_{\rm X}}}
\def\mmaj{\ifmath{M_{\rm maj}}}
\def\sM{\ifmath{{\bf M}}}
\def\BM{\ifmath{{\bf B_M}}}
\def\mLs{\ifmath{{{\bf m}_L^2}}}
\def\mes{\ifmath{{{\bf m}_e^2}}}
\def\mns{\ifmath{{{\bf m}_\nu^2}}}
\def\mQs{\ifmath{{{\bf m}_Q^2}}}
\def\mus{\ifmath{{{\bf m}_u^2}}}
\def\mds{\ifmath{{{\bf m}_d^2}}}
\def\mHus{\ifmath{m_{H_u}^2}}
\def\mHds{\ifmath{m_{H_d}^2}}
\def\Ann{\ifmath{A_\nu}}
\def\Aen{\ifmath{A_e}}
\def\Y{\ifmath{{\bf Y}}}
\def\Yx{\ifmath{{{\bf Y}_x}}}
\def\Yn{\ifmath{{{\bf Y}_\nu}}}
\def\Ye{\ifmath{{{\bf Y}_e}}}
\def\Yu{\ifmath{{{\bf Y}_u}}}
\def\Yd{\ifmath{{{\bf Y}_d}}}

\def\Ax{\ifmath{{{\bf A}_x}}}
\def\An{\ifmath{{{\bf A}_\nu}}}
\def\Ae{\ifmath{{{\bf A}_e}}}
\def\Au{\ifmath{{{\bf A}_u}}}
\def\Ad{\ifmath{{{\bf A}_d}}}

\def\fr{\ifmath{\frac{1}{16 \pi^2}}}
\def\frsq{\ifmath{\frac{1}{(16 \pi^2)^2}}}
\def\lg{\ifmath{\log \!\!\left(\!\frac{\mgut}{\mmaj}\!\right)}}
\def\dag{\ifmath{{\!\!\!\dagger}}}
\def\tra{\ifmath{{\!\!\!T}}}
\def\con{\ifmath{{\!\!\!*}}}
\def\YYn{\ifmath{\Yn^{\dag} \Yn}}
\def\YYnb#1{\ifmath{( \Yn^{\dag} \Yn )_{#1}}}
\def\Tr{\ifmath{{\rm Tr}}}
\def\beto{\ifmath{\beta^{(1)}}}
\def\bett{\ifmath{\beta^{(2)}}}

\def\sss{{\cal S}^\prime}
\def\trym{{\cal S}}

\def\boxeq#1{\boxit{${\displaystyle #1 }$}}          

\begin{document}
\begin{titlepage}

\title{ \bf{
\mbox{ \hspace{-1.2cm} 
Constraints on the rare tau decays from $\mu\rightarrow e\gamma$}
in the supersymmetric see-saw model}}
\vskip3in \author{{\bf Alejandro Ibarra$^1$} and
{\bf Cristoforo Simonetto$^2$
\footnote{\baselineskip=16pt {\small E-mail addresses: {\tt
alejandro.ibarra@desy.de, cristoforo.simonetto@ph.tum.de}}}}
\hspace{3cm}\\
{\small $^1$ DESY,  Theory Group, Notkestrasse 85, 22603 Hamburg, Germany}\\
{\small $^2$ Physik Department T30, Technische Universit\"at M\"unchen,}\\[-0.05cm]
{\it\small James-Franck-Strasse, 85748 Garching, Germany}.
}  \date{}  \maketitle  \def\baselinestretch{1.15}
\begin{abstract}
\noindent 
It is now a firmly established fact that all family lepton
numbers are violated in Nature. In this paper we discuss
the implications of this observation for future searches
for rare tau decays in the supersymmetric see-saw model.
Using the two loop renormalization group evolution of the soft
terms and the Yukawa couplings
we show that there exists a lower
bound on the rate of the rare process $\mu \rightarrow e\gamma$
of the form $BR(\mu\rightarrow e\gamma)\gsim
C \times BR(\tau\rightarrow \mu\gamma)BR(\tau\rightarrow e\gamma)$,
where $C$ is a constant that depends on supersymmetric parameters.
Our only assumption is the absence of cancellations among
the high-energy see-saw parameters.
We also discuss the implications of this bound for future searches
for rare tau decays. In particular, for large regions
of the mSUGRA parameter space, we show that present $B$-factories
could discover either $\tau\rightarrow \mu\gamma$ or
$\tau\rightarrow e\gamma$, but not both. 
\end{abstract}

\thispagestyle{empty}
\vspace*{0.2cm} \leftline{February 2008} \leftline{}

\vskip-16.5cm \rightline{DESY 08-018}
\rightline{TUM-HEP-682/08}
\vskip3in

\end{titlepage}
\setcounter{footnote}{0} \setcounter{page}{1}
\newpage
\baselineskip=20pt

\noindent

\section{Introduction}

The renormalizable part of the Standard Model Lagrangian 
is invariant under four global $U(1)$ symmetries,
namely baryon number, $B$,  and the three family lepton numbers, 
$L_e$, $L_\mu$ and $L_\tau$.
This invariance has been for many years considered accidental
and expected to be broken by additional terms in the Lagrangian, 
possibly non-renormalizable. 
Whereas experiments searching for proton decay have not 
provided yet any evidence for baryon 
number violation, it is nowadays a firmly established experimental
fact that the three family lepton numbers are violated in the 
neutrino sector. Namely, the disappearance of electron neutrinos 
on their way from the Sun indicated by the Homestake chlorine 
detector~\cite{Davis:1968cp}, SAGE~\cite{Abdurashitov:2002nt}, 
GALLEX/GNO~\cite{Hampel:1998xg,Altmann:2005ix},
Kamiokande~\cite{Fukuda:1996sz}, Super-Kamiokande~\cite{Smy:2003jf}
and Borexino~\cite{Collaboration:2007xf}, 
and unequivocally confirmed by SNO~\cite{SNO}, 
proves the violation of $L_e$. This is further supported by
the disappearance of electron antineutrinos observed
by the reactor experiment KamLAND~\cite{Eguchi:2002dm}. On the other hand, 
the atmospheric neutrino anomaly discovered by Kamiokande and 
IMB~\cite{Becker-Szendy:1992hq},
and explained by Super-Kamiokande~\cite{Fukuda:1998mi}, 
Soudan2~\cite{Sanchez:2003rb} and MACRO~\cite{Ambrosio:2001je} 
as an oscillation of muon neutrinos into a different neutrino species, 
proves the violation of $L_\mu$.
The disappearance of muon neutrinos reported by
the long baseline accelerator experiments K2K~\cite{Aliu:2004sq} 
and MINOS~\cite{Michael:2006rx} supports this conclusion. Finally,
the observation of tau neutrino appearance in the
atmospheric neutrino flux by Super-Kamiokande~\cite{Abe:2006fu}
indicates the violation of $L_\tau$.

The most economical way to accommodate the family lepton number
violation in the Standard Model is by adding to the leptonic
Lagrangian a dimension five operator~\cite{Weinberg:1979sa}
\begin{equation}
{\cal L}_{\rm lep}=-{e^c_R}_i \Ye_{ij} L_j H^* 
-\frac{\alpha_{ij}}{\Lambda} (L_i H)(L_j H)
+{\rm h.c.}\;,
\label{dimension5}
\end{equation}
where $L_i$ and $e^c_R$ are the left-handed and right-handed 
leptons respectively, $i,j=1,2,3$, $H$ is the Higgs doublet
and $\Lambda$ a mass parameter. In fact, this operator not only
violates the three family lepton numbers but also the total lepton
number. After the electroweak symmetry breaking, a Majorana mass term
is generated for the left-handed neutrinos, 
${\cal M}_{ij}=\frac{\alpha_{ij}}{\Lambda} \langle H^0 \rangle^2$,
being the smallness of neutrino masses attributed to 
small values of the couplings $\alpha_{ij}$ and/or to a large 
value of $\Lambda$. 

In this minimal framework, lepton flavour
violation is generated by the same \mbox{operator} that generates
neutrino masses. Therefore, the rate of any lepton flavour violating 
process will be proportional to the neutrino masses. Indeed, a
detailed calculation shows that~\cite{meg_SM}:
\begin{equation}
BR(l_j\rightarrow l_i \gamma)=\frac{3\alpha}{32\pi}
\Bigg|\frac{\Delta m_{\rm sol}^2}{M_W^2}U^*_{i2}U_{j2}
+\frac{\Delta m_{\rm atm}^2}{M_W^2}U^*_{i3}U_{j3}\Bigg|^2
BR(l_j\rightarrow l_i \nu_j \bar \nu_i)\;,
\end{equation}
which gives $BR(\tau\rightarrow \mu \gamma)\sim 10^{-54}$,
$BR(\mu\rightarrow e \gamma)\sim10^{-57}$, 
$BR(\tau\rightarrow e \gamma)\sim 10^{-57}$,
far below the sensitivity of present and projected experiments
(see Table 1)\footnote{In the Table we restrict ourselves to
bounds that were published by the time of writing this
paper. We note however that the Belle Collaboration has reported 
the bounds $BR(\tau\rightarrow \mu\gamma)<4.5 \times 10^{-8}$
and $BR(\tau\rightarrow e\gamma)<1.2 \times 10^{-7}$ in the yet 
unpublished preprint~\cite{Hayasaka:2007vc}.}. 

\begin{table}[t]
\begin{center}
\begin{tabular}{|c|c|c|}
\hline 
& present bound & projected bound \\\hline
$BR(\mu\rightarrow e\gamma)$ & $1.2\times 10^{-11}$~\cite{Brooks:1999pu}&
$10^{-13}$~\cite{MEG} \\
$BR(\tau\rightarrow e\gamma)$ & $1.1\times 10^{-7}$~\cite{Aubert:2005wa}& 
$10^{-9}$~\cite{Akeroyd:2004mj} \\
$BR(\tau\rightarrow \mu\gamma)$ & $6.8\times 10^{-8}$~\cite{Aubert:2005ye}& 
$10^{-9}$~\cite{Akeroyd:2004mj} \\
\hline
\end{tabular}
\end{center}
\caption{\small Present and projected bounds on the rare lepton decays.}
\end{table}

Nevertheless, the Lagrangian Eq.~(\ref{dimension5}) describes 
just an effective theory and new degrees of freedom are expected 
to arise above the scale $\Lambda$. The interactions
of the new degrees of freedom with the lepton fields
are likely to contain additional sources of flavour violation
that can enhance the branching ratios of the rare decays by 
many orders of magnitude, bringing them to the reach of 
future experiments. For this reason, rare lepton decays are 
considered very powerful probes for physics beyond the Standard Model. 

The supersymmetric (SUSY) see-saw mechanism is probably the best
motivated high energy theory to generate small neutrino 
masses~\cite{seesaw} . In this framework the particle content of
the Minimal Supersymmetric Standard Model (MSSM) is extended
with three right-handed neutrino superfields, ${\nu_{R}}_i$, $i=1,2,3$,
singlets under the Standard Model gauge group.
Imposing $R$-parity conservation, the leptonic superpotential reads:
\begin{equation}
W_{\rm lep}= {e_R^c}_i \Ye_{ij} L_j  H_d+
{\nu_R^c}_i \Yn_{ij} L_j  H_u
- \frac{1}{2}{\nu_R^c}_i{\sM}_{ij}{\nu_R^c}_j \;,
\label{superp} 
\end{equation}
where $H_u$ and $H_d$ are the hypercharge $+1/2$ and $-1/2$ Higgs doublets,
respectively, $\Ye$ and $\Yn$ are the matrices of charged lepton and 
neutrino Yukawa couplings, respectively, 
and  $\sM$ is a $3\times 3$ Majorana mass matrix. 
It is natural to assume that the overall scale of $\sM$, 
which we will denote by $\mmaj$, is much larger than the 
electroweak scale or any soft mass. If this is the case, 
at energies below $\mmaj$ the theory is well described by the following 
effective superpotential:
\begin{equation} 
W^{\rm eff}_{\rm lep}={e_R^c}_i \Ye_{ij} L_j  H_d+\frac{1}{2}
\left(\Yn^\tra{\sM}^{-1}\Yn \right)_{ij} (L_i H_u)(L_j H_u)\;,
\label{effsuperp}
\end{equation}
that generates the fermionic Lagrangian Eq.~(\ref{dimension5}).
In the phenomenological studies it is convenient to work in the
leptonic basis where the charged lepton
Yukawa coupling is diagonal, $\Ye={\rm diag}(y_e, y_\mu, y_\tau)$.
Then, in this basis, the neutrino mass matrix is given by
\begin{equation}
{\cal M}=\left(\Yn^\tra {\sM}^{-1}\Yn\right) 
\langle H_u^0\rangle^2 \;,
\end{equation}
whose eigenvalues are naturally very small due to
the suppression by the large right-handed neutrino mass scale.

From the point of view of generating small neutrino masses,
the non-supersymmetric version of the see-saw mechanism is 
equally natural. Nonetheless, in the non-super\-symmetric 
see-saw model the presence of very heavy new degrees of freedom
interacting with the Higgs doublet introduces a serious naturalness
problem. The Higgs mass acquires quadratically-divergent radiative
corrections which would drive the Higgs mass to values of the
order of the Majorana mass scale~\cite{hierarchy}.
However, in the supersymmetric version this divergence
is automatically canceled by the presence of right-handed
sneutrinos with a mass essentially identical to the mass
of their corresponding right-handed neutrinos. Therefore, the 
supersymmetric see-saw mechanism can accommodate simultaneously 
tiny neutrino masses and a relatively light electroweak scale without
serious fine-tunings.

In addition to the flavour violation stemming from the right-handed
neutrino Yukawa couplings, the supersymmetric see-saw model
contains additional sources of lepton flavour violation in 
the soft SUSY breaking Lagrangian~\cite{offdiag}: 
\begin{eqnarray}  
-{\cal L}_{\rm soft}&=&\ 
(\mLs)_{ij} \widetilde L^*_i  \widetilde L_j\ +
(\mes)_{ij} \widetilde e^*_{Ri}  \widetilde e_{Rj}\ + 
(\mns)_{ij} \widetilde \nu^*_{Ri}  \widetilde \nu_{Rj}\ +  
\nonumber \\
&&\ \left(\Ae_{ij} \widetilde e^*_{Ri} H_d \widetilde L_j +
\An_{ij} \widetilde \nu^*_{Ri} H_u \widetilde L_j +
{\rm h.c.}\right)+ 
{\rm etc}\;.
\end{eqnarray}
where $\widetilde L_i$, $\widetilde e_{Ri}$ and $\widetilde \nu_{Ri}$ 
are the supersymmetric partners of the left-handed lepton doublets, 
right-handed charged leptons and right-handed neutrinos, respectively,
$\mLs$, $\mes$ and $\mns$ are their corresponding soft mass matrices squared,
and $\Ae$ and $\An$ are the charged lepton and neutrino soft trilinear
terms. 

The flavour violation in the slepton sector contributes
through one loop diagrams to different flavour violating processes
such as rare muon and tau decays, $K^0_L \rightarrow e^\pm \mu^\mp$ or 
$\mu-e$ conversion in nuclei. The strong bounds on these  
processes restrict very severely the structure
of the soft mass matrices at low energies~\cite{FCNC}, suggesting an
approximately flavour universal structure:
$(\mLs)_{ij}\simeq m_L^2 \delta_{ij}$, 
$(\mes)_{ij}\simeq m_e^2 \delta_{ij}$, 
$(\Ae)_{ij}\simeq {A_e}{\Ye}_{ij}$.
The most plausible explanation for this structure 
is to assume that supersymmetry is broken in a hidden
sector and the breaking is transmitted to the visible sector
by a flavour blind mediation mechanism. If this is the case,
the soft terms would be strictly flavour universal at some high energy
scale: 
\begin{eqnarray}
&&(\mLs)_{ij}=m_L^2 \delta_{ij}, ~~~~
(\mes)_{ij}= m_e^2 \delta_{ij}, ~~~~
(\mns)_{ij}= m_{\nu}^2 \delta_{ij}, \nonumber \\
&&\hspace{1.5cm}(\Ae)_{ij}= \Aen\;{\Ye}_{ij}, ~~~~
(\An)_{ij}= \Ann\;{\Yn}_{ij}\;.
\end{eqnarray}

Interestingly, if this high energy scale is larger than the 
right-handed neutrino masses, the flavour violation in the
neutrino Yukawa couplings will propagate through radiative effects
to the soft terms~\cite{Borzumati:1986qx}. As a consequence, even 
under the most conservative assumption for the soft terms from
the point of view of lepton flavour violation, in the
supersymmetric see-saw model some amount of flavour violation 
is always expected at low energies.

The discovery of small neutrino masses as a hint of
the see-saw mechanism and the continuous improvement
in sensitivity of the experiments searching for rare lepton 
decays have stimulated in recent years the interest 
in the radiative generation of lepton flavour violation.
Different groups have estimated the predictions for 
various lepton flavour violating processes in 
some specific high-energy frameworks based on Grand 
Unification~\cite{GUTS}, flavour models~\cite{flavourmodels},
texture zeros~\cite{texturezeros}
or string theory~\cite{strings}, but also pursuing a more
phenomenological approach, where the predictions are 
somehow connected to low energy 
observables~\cite{Hisano:1995cp,Hisano:1998fj,Casas:2001sr,general,Petcov:2003zb}. 
A general expectation of all these analyses is that
the observation of rare lepton decays could be possible
at the next round of experiments.
Nevertheless, it should be stressed that there are many
optimistic assumptions underlying this conclusion,
and the non-observation of rare decays in future experiments
would by no means exclude the supersymmetric see-saw model.

In this paper we will adopt a completely phenomenological 
approach and we will derive the relation 
among the branching ratios $BR(\mu\rightarrow e\gamma)\gsim
C \times BR(\tau\rightarrow \mu\gamma)BR(\tau\rightarrow e\gamma)$
that holds independently of the high-energy see-saw
parameters, with the only assumption of the absence
of cancellations. In this relation, however, some model 
dependence will remain associated to our ignorance on the 
supersymmetric parameters, which is encoded in the constant $C$. 
To derive our main result we will
present in Section 2 the scenario that saturates the bound and
yields the minimal rate for $\mu\rightarrow e\gamma$. 
In Section 3 we will compute the off-diagonal terms
of the slepton mass matrices in this extremal scenario taking
into account the two loop renormalization group equations
and we will derive relations among them. Using these
relations, we will derive in Section 4 the lower bound
on $BR(\mu\rightarrow e\gamma)$ in terms of the branching ratios of
the rare tau decays. We will also discuss the implications of this
bound for present and future searches for rare tau decays
for different supersymmetric benchmark points. Finally, 
in Section 5 we will present our conclusions. We will
also present an Appendix containing the complete two loop renormalization
group equations for the soft SUSY breaking terms, including the effects 
of the neutrino Yukawa couplings, which to the best of our knowledge
have not been given explicitely in the literature.

\section{Scenarios with minimal rates for $\mu\rightarrow e\gamma$}

In the MSSM extended with right-handed neutrinos there are sources 
of lepton flavour violation both in the soft SUSY breaking Lagrangian 
and in the SUSY conserving Lagrangian. The latter can be 
completely encoded in the neutrino Yukawa couplings by means
of a basis transformation. Neutrino oscillations 
require the presence of lepton flavour violation in the
neutrino Yukawa couplings, but not in the soft SUSY breaking
Lagrangian. Therefore, the minimal rate for $\mu\rightarrow e\gamma$
will clearly occur in scenarios where there is no lepton flavour violation
in the soft SUSY breaking Lagrangian. There are in fact some
well motivated supersymmetric scenarios where the soft breaking
parameters are flavour universal at some high energy scale.
However, if this scale is larger than the 
mass of the right-handed neutrinos\footnote{This is the
case for example in scenarios with minimal supergravity,
dilaton-dominated SUSY breaking, or gauge mediation where
the masses of the messenger particles are larger than the
masses of the right-handed neutrinos.}, the flavour violation in 
the supersymmetric part of the Lagrangian will propagate
to the soft SUSY breaking terms through quantum effects.

For generic neutrino Yukawa couplings, the off-diagonal
elements of the soft SUSY breaking terms read
at low energies, in the leading-log approximation,
\begin{eqnarray}
\left(\mLs\right)_{ij} & \simeq & -\frac{1}{8\pi^2}
(m_L^2+m_\nu^2+\mHus+|\Ann|^2)(\YYn)_{ij} 
\log\left(\frac{\mgut}{\mmaj}\right)\ , \nonumber\\
\left(\mes\right)_{ij} & \simeq & 0\ , \nonumber\\  
(\Ae)_{ij}&\simeq &   \frac{-1}{8\pi^2} (2\Ann+A_e) \Ye (\YYn)_{ij}
 \log\left(\frac{\mgut}{\mmaj}\right)\;, 
\label{softafterRG} 
\end{eqnarray}
where $i\neq j$ and $\mgut$ is some high energy scale that we identify 
with the Grand Unification scale, $\mgut =2\times 10^{16}$ GeV. 
The size of the off-diagonal elements depends very strongly
on the flavour structure of the neutrino Yukawa couplings, 
which in the see-saw model is not directly connected to the
flavour structure of the low energy neutrino mass matrix. 
In fact, there is an infinite set of 
neutrino Yukawa couplings compatible with a given set
of low energy data~\cite{Casas:2001sr}. Among all those Yukawa couplings,
the minimal rate for $\mu\rightarrow e\gamma$ will clearly occur 
in the scenario where
\begin{equation}
\left(\YYn\right)_{21}(\mmaj)
=y_{12}^* y_{11} +y_{22}^* y_{21}+y_{32}^* y_{31}=0\,.
\label{YYd12eq0}
\end{equation}

Assuming that there are no cancellations among the different terms,
this condition is satisfied in the following eight situations:
\begin{eqnarray}
&(a)~~~   y_{11}=0 ,   y_{21}=0 ,   y_{31}=0\;, ~~~~~~~
(e)~~~   y_{12}=0 ,   y_{21}=0 ,   y_{31}=0 \;, \nonumber \\
&(b)~~~   y_{11}=0 ,   y_{21}=0 ,   y_{32}=0\;, ~~~~~~~
(f)~~~   y_{12}=0 ,   y_{21}=0 ,   y_{32}=0\;,  \nonumber \\
&(c)~~~   y_{11}=0 ,   y_{22}=0 ,   y_{31}=0\;, ~~~~~~~
(g)~~~   y_{12}=0 ,   y_{22}=0 ,   y_{31}=0 \;, \nonumber \\
&(d)~~~   y_{11}=0 ,   y_{22}=0 ,   y_{32}=0 \;,~~~~~~~
(h)~~~   y_{12}=0 ,   y_{22}=0 ,   y_{32}=0 \;, 
\label{cases}
\end{eqnarray}
where the Yukawa matrix elements are evaluated at $\mmaj$.
The cases $(a)$ and $(h)$ preserve electron and muon lepton numbers
respectively. Since in this basis the neutrino Yukawa coupling
is the only source of lepton flavour violation, the complete
Lagrangian will preserve at least one family lepton number. 
This invariance is inherited by the effective theory, 
in conflict with the flavour conversions observed in neutrino oscillations. 
On the other hand, the remaining six possibilities $(b)-(g)$ 
violate all family lepton numbers and
lead to a neutrino mass matrix that schematically reads
\begin{equation}
{\cal M}\sim\begin{pmatrix} \times  & 0 & \times \\
                             0 & \times &\times \\ 
                           \times & \times &\times \end{pmatrix}
\end{equation}
and which is allowed by present experiments (this mass matrix leads to the
prediction $\sin\theta_{13}\simeq\frac{1}{2}\sqrt{\frac{\Delta m^2_{\rm sol}}
{\Delta m^2_{\rm atm}}}$ $\sin2\theta_{\rm sol}
\simeq 0.08$ for the case with normal
neutrino mass hierarchy and $\sin\theta_{13}\simeq\frac{1}{4} 
\sqrt{\frac{\Delta m^2_{\rm sol}}
{\Delta m^2_{\rm atm}}}$ $\sin2\theta_{\rm sol}
\simeq 0.04$ for the case with inverted neutrino mass hierarchy).
Therefore, 
from Eq.~(\ref{cases}) it is straightforward to check
that the experimentally allowed neutrino Yukawa textures which 
lead to a vanishing rate for $\mu\rightarrow e\gamma$
necessarily lead to $(\YYn)_{31}(\mmaj)\neq 0$ and
$(\YYn)_{32}(\mmaj)\neq 0$, unless unnatural cancellations 
in Eq.~(\ref{YYd12eq0}) are taking place.

The presence of exact zeros in the Yukawa matrices as
in Eq.~(\ref{cases}) can be justified by symmetries. 
However, symmetries are not expected to hold at the decoupling scale 
but at the Grand Unification scale, $\mgut$. Again, and barring
cancellations, the minimal rate for $\mu\rightarrow e\gamma$
will occur when $(\YYn)_{21}(\mgut)= 0$, that as was argued above
necessarily implies $(\YYn)_{31}(\mgut)\neq 0$ and 
$(\YYn)_{32}(\mgut)\neq 0$. 

The key point of this paper is that even when the soft terms are 
flavour universal at $\mgut$ and $(\YYn)_{21}(\mgut)=0$,
the flavour violation in the $\tau-\mu$ and the $\tau-e$ sectors, 
will propagate through radiative corrections to the $\mu-e$ sector, 
inducing small though non-vanishing values for $(\mLs)_{21}$
and $(\Ae)_{12,21}$ at low energies. We will show in this
paper that in this very special case the rate of
$\mu\rightarrow e\gamma$ is related to the rates of the
rare tau decays through
\begin{equation}
BR(\mu\rightarrow e\gamma)\simeq C \times BR(\tau\rightarrow \mu\gamma)
BR(\tau\rightarrow e\gamma)\;,
\end{equation}
where the proportionality constant, $C$, will be determined in 
the next sections. 

This value is attained by the
experimentally allowed neutrino Yukawa textures satisfying
$(\YYn)_{21}(\mgut)=0$, which is the scenario producing the 
minimal rate for $\mu\rightarrow e\gamma$.
Therefore, and barring cancellations, for any other Yukawa 
coupling compatible with the low energy neutrino parameters
the following bound will hold:
\begin{equation}
\boxeq{
BR(\mu\rightarrow e\gamma)\gsim C\times BR(\tau\rightarrow \mu\gamma)
BR(\tau\rightarrow e\gamma)}\;.
\label{mainresult}
\end{equation}
This inequality is the main result of this paper. As we will 
see later, this bound has important implications
for the searches for rare tau decays in present and future experiments.

To finish this section we would like to comment on 
the possibility sometimes discussed in the literature
of establishing model independent correlations between 
the branching ratios of the rare decays of the form
$BR(\tau\rightarrow \mu\gamma)/BR(\mu\rightarrow e\gamma)
\simeq {\rm constant}$.
We would like to clarify this point stressing that in a general see-saw 
model with three right-handed neutrinos such correlations {\it cannot}
be established on model independent grounds. 

It can be proved
that there is a one to one mapping between the high energy
see-saw parameters, $\Yn$ and $\sM$, and the matrices 
${\cal M}$ and $\YYn$~\cite{Davidson:2001zk}. The former is the 
neutrino mass matrix and the latter, the matrix that participates
in the radiative corrections to the soft SUSY breaking terms,
which is in principle measurable at low energies once the boundary
conditions for the soft terms have been specified. In particular,
the off-diagonal elements of $\YYn$ are directly related to the 
branching ratios of the rare decays. Being this mapping
bijective, one could use as parameters
of the see-saw model either the familiar set $\{{\Yn},{\sM}\}$
(the top-down parametrization) or the less familiar one 
$\{{\cal M},\YYn\}$ (the bottom-up parametrization). Clearly, in 
the bottom-up parametrization the branching ratios for the
rare decays are inputs, and as such are completely uncorrelated.
For the same reason, it is not possible to establish on model
independent grounds correlations between the branching ratios
of the rare decays and neutrino parameters, such as $\sin\theta_{13}$, the
neutrino mass spectrum or the CP violating phases. 
This result also holds for the most
restricted case with just two right-handed neutrinos~\cite{Ibarra:2005qi}. 

A more explicit proof of this result can be derived as follows
using the familiar top-down parametrization. 
Let us first write the effective neutrino mass
matrix in terms of the high-energy parameters in the basis where
the charged lepton Yukawa coupling and the right-handed mass
matrix are simultaneously diagonal:
\begin{equation}
{\cal M}_{ij}=\left(\frac{y_{1i} y_{1j}}{M_1}+
\frac{y_{2i} y_{2j}}{M_2}+\frac{y_{3i} y_{3j}}{M_3}\right)
\langle H_u^0\rangle^2 \;.
\end{equation}
One could conceive a scenario where $\frac{y_{3i} y_{3j}}{M_3}\ll
\frac{y_{1i} y_{1j}}{M_1},\frac{y_{2i} y_{2j}}{M_2}$ for all $i,j=1,2,3$.
This could occur for instance 
when the mass of the heaviest right handed neutrino
mass is much larger than the masses of the other two right-handed
neutrinos. If this is the case, the third row of the neutrino
Yukawa matrix is completely unconstrained from neutrino observations.
In addition, it could occur that $y_{3i}$ are much larger 
than $y_{1i}, y_{2i}$, implying that 
$(\YYn)_{ij}\simeq y^*_{3i} y_{3j}$. Therefore,
the branching ratios for the rare decays would be essentially 
determined by the parameters $y_{3i}$, that are
completely unconstrained from low energy data. As a 
result, the ratio $BR(\tau\rightarrow \mu\gamma)/ BR(\mu\rightarrow e\gamma)$
can range from zero (when $y_{33}=0$, $y_{31},y_{32}\neq 0$)
to infinity (when $y_{31}=0$, $y_{32},y_{33}\neq 0$), being both
situations compatible with neutrino observations.
Additionally, since the parameters that determine the branching
ratios, $y_{3i}$, are unconstrained from low energy data, 
no model independent correlation can be established between
rare decays and neutrino parameters. 

In order to establish correlations between
two branching ratios, or correlations between one branching ratio
and low energy neutrino parameters, it is {\it necessary}
to make assumptions about the high-energy see-saw parameters, 
for instance motivated by  Grand Unified theories 
or by flavour models. 
In this paper we make what we consider the most minimal 
assumption about the high-energy see-saw parameters, namely that 
in constructing the derived quantities ${\cal M}$ and $\YYn$
no cancellations occur. Then, the no-go theorem presented 
in~\cite{Davidson:2001zk,Ibarra:2005qi} can be circumvented,
opening the possibility of deriving relations among the branching 
ratios. Although the relation that results, 
$BR(\mu\rightarrow e\gamma)\gsim C\times BR(\tau\rightarrow \mu\gamma)
BR(\tau\rightarrow e\gamma)$, is not
model independent in the strict sense, it is valid in a
very large class of models, characterized by the absence
of unnatural cancellations. In addition, this bound will 
proof to have remarkably strong implications given the 
generality of our assumptions, as we will see in Section 4.

\section{Radiative generation of lepton flavour violation in
the soft mass matrices}

The Renormalization Group Equations (RGEs) for the relevant
soft terms read
\begin{eqnarray}
\frac{d\mLs}{dt}&=&\fr \beta_\mLs^{(1)} +\frsq \beta_\mLs^{(2)}+...\;, \\
\frac{d\mes}{dt}&=&\fr \beta_\mes^{(1)} +\frsq \beta_\mes^{(2)}+...\;, \\
\frac{d\Ae}{dt}&=&\fr \beta_\Ae^{(1)} +\frsq \beta_\Ae^{(2)}+...\;,
\end{eqnarray}
where $\beta^{(1)}$ and $\beta^{(2)}$ indicate respectively
the one and two loop $\beta$-functions. The complete expression
for the $\beta$-functions of the soft terms in the supersymmetric 
see-saw model can be found in the Appendix.   

The solution to the RGEs can be well approximated by the trapezium 
rule, which keeping just the one and two loop contributions reads:
\begin{eqnarray}
\mLs(\mmaj)&\simeq& \mLs(\mgut)-
\fr \frac{1}{2}\left[\beta_\mLs^{(1)}(\mgut)+\beta_\mLs^{(1)}(\mmaj)\right] 
\log\left(\frac{\mgut}{\mmaj}\right)
\nonumber \\
&& -\frsq \frac{1}{2}\left[\beta_\mLs^{(2)}(\mgut)+\beta_\mLs^{(2)}(\mmaj)\right]
\log\left(\frac{\mgut}{\mmaj}\right)\;,  \label{approx-mL2}\\
\mes(\mmaj)&\simeq& \mes(\mgut)-
\fr \frac{1}{2}\left[\beta_\mes^{(1)}(\mgut)+\beta_\mes^{(1)}(\mmaj)\right] 
\log\left(\frac{\mgut}{\mmaj}\right)
\nonumber \\
&& -\frsq \frac{1}{2}\left[\beta_\mes^{(2)}(\mgut)+\beta_\mes^{(2)}(\mmaj)\right]
\log\left(\frac{\mgut}{\mmaj}\right) \;, \label{approx-me2}\\
\Ae(\mmaj)&\simeq& \Ae(\mgut)-
\fr \frac{1}{2}\left[\beta_\Ae^{(1)}(\mgut)+\beta_\Ae^{(1)}(\mmaj)\right] 
\log\left(\frac{\mgut}{\mmaj}\right)
\nonumber \\
&& -\frsq \frac{1}{2}\left[\beta_\Ae^{(2)}(\mgut)+\beta_\Ae^{(2)}(\mmaj)\right]
\log\left(\frac{\mgut}{\mmaj}\right)\;. \label{approx-Ae}
\end{eqnarray}

For the Yukawa couplings leading to a minimal rate for
$\mu\rightarrow e\gamma$, Eq.~(\ref{cases}),
the 31 and 32 entries in the left-handed slepton mass matrix are generated 
at order ${\cal O}(Y^2_{\nu})$. Therefore, noting that 
$\beta_\mLs^{(1)}(\mmaj)=\beta_\mLs^{(1)}(\mgut)+{\cal O}(Y^4_{\nu})$
and that the two loop $\beta$-functions are ${\cal O}(Y^4_{\nu})$,
it follows from Eq.~(\ref{approx-mL2}) that
\begin{equation}
\mLs(\mmaj)_{31,32}= - \fr (\beta_\mLs^{(1)})_{31,32}(\mgut) \lg
+{\cal O}(Y^4_{\nu})\;.
\label{approx-mL2-13}
\end{equation}
Using Eq.~(\ref{beta1loop-mLs}) in the Appendix 
for the one loop $\beta$-function we find
\begin{equation}
\mLs(\mmaj)_{31,32}= -\frac{1}{8\pi^2}
(m_L^2+m_\nu^2+\mHus+|\Ann|^2) \YYnb{31,32}\lg+{\cal O}(Y^4_{\nu})\;,
\end{equation}
where all the soft terms in the right-hand side of this equation
and $\YYnb{31,32}$ are understood to be evaluated at the scale $M_X$.

On the other hand, since $\YYnb{21}(\mgut)=0$ the previous equation
indicates that $\mLs(\mmaj)_{21}$ vanishes at order 
${\cal O}(Y^2_{\nu})$ and is only generated at higher order in
perturbation theory. Keeping only terms of ${\cal O}(Y^4_{\nu})$,
it follows that 
$(\beta_\mLs^{(2)})_{21}(\mmaj)=(\beta_\mLs^{(2)})_{21}(\mgut)+
{\cal O}(Y^6_{\nu})$ and $(\beta_\mLs^{(1)})_{21}(\mgut)=0$. Therefore,
\begin{equation}
(\mLs)_{21}(\mmaj)= 
-\fr \frac{1}{2} (\beta_\mLs^{(1)})_{21}(\mmaj)\lg
 -\frsq (\beta_\mLs^{(2)})_{21}(\mgut) \lg
+{\cal O}(Y^6_{\nu})\;.
\label{approx-mL2-12}
\end{equation}
The one loop $\beta$-function at the Majorana mass scale
is proportional to $\YYnb{21}(\mmaj)$, that following
Eq.~(\ref{beta1loopYn}) in the Appendix is generated
radiatively by the effect of $\YYnb{32}$ and $\YYnb{31}$.
The result is:
\begin{equation}
\YYnb{21}(\mmaj)= -\frac{3}{8\pi^2}\YYnb{32}^* \YYnb{31}\lg
+{\cal O}(Y^6_{\nu})\;.
\label{yyd12}
\end{equation}

Using Eqs.~(\ref{approx-mL2-12},\ref{yyd12}) and 
Eqs.~(\ref{beta1loop-mLs},\ref{beta2loop-mLs}) in the Appendix, 
it is straightforward to compute $(\mLs)_{21}(\mmaj)$. The result is:
\begin{eqnarray} 
(\mLs)_{21}(\mmaj)&=&  \frsq (m_L^2+\mHus+m_\nu^2+2 |\Ann|^2)
 \nonumber \\
&& \hspace{-2cm} 
 \times\left[12 \lg^2 +8 \lg\right]\YYnb{32}^* \YYnb{31}+{\cal O}(Y^6_{\nu})\;.
\end{eqnarray}
Recall from the previous section that when $\YYnb{21}(\mgut)=0$, 
the observed pattern of neutrino mixing angles requires 
$\YYnb{31}(\mgut)\neq 0$ and $\YYnb{32}(\mgut)\neq 0$.
Then, $(\mLs)_{21}(\mmaj)$ will always be generated {\it at least} at order
${\cal O}(Y^4_{\nu})$: it will be generated at order ${\cal O}(Y^2_{\nu})$
when $\YYnb{21}(\mgut)$ is different from zero and at order 
${\cal O}(Y^4_{\nu})$ when it is equal to zero. Therefore, and barring
cancellations, the observed violation of all family lepton numbers
in neutrino oscillations {\it necessarily} implies in the 
supersymmetric see-saw model a contribution to the process 
$\mu\rightarrow e\gamma$ from the off-diagonal entries of the soft 
terms.

Following the same steps , one can calculate the off-diagonal elements
of the trilinear soft terms. For the elements generated at 
${\cal O}(Y^2_{\nu})$ we obtain
\begin{eqnarray}
 (\Ae)_{31,32} &=& - \fr \lg (2\Ann + \Aen) y_\tau \YYnb{31,32}
+{\cal O}(Y^4_{\nu})\;,\\
 (\Ae)_{13,23} &=& - \fr \lg (2\Ann + \Aen) y_{e,\mu} \YYnb{31,32}^*
+{\cal O}(Y^4_{\nu})\;,
\end{eqnarray}
and for the elements generated at ${\cal O}(Y^4_{\nu})$,
\begin{eqnarray}
(\Ae)_{21} &=& \left(\fr \right)^2 
\left[(14 \Ann +\frac{7}{2}\Aen) \lg^2+(8\Ann+2\Aen)\lg\right] \nonumber \\
&&\times y_\mu \YYnb{32}^* \YYnb{31}+{\cal O}(Y^6_{\nu})\;, \\
(\Ae)_{12} &=& \left(\fr \right)^2 
\left[(14 \Ann +\frac{7}{2}\Aen) \lg^2+(8\Ann+2\Aen)\lg\right] \nonumber \\
&&\times y_e \YYnb{31}^* \YYnb{32}+{\cal O}(Y^6_{\nu})\;.
\end{eqnarray}

Finally, there are additional sources of lepton flavour violation
stemming from the radiatively generated off-diagonal
entries in the charged lepton Yukawa coupling.
\mbox{Analogously} to the previous discussion, $(\Ye)(\mmaj)_{12}$
and $(\Ye)(\mmaj)_{21}$ get values of ${\cal O}(Y^4_{\nu})$,
while the remaining off-diagonal terms get values 
of ${\cal O}(Y^2_{\nu})$. Rotating the leptonic fields
to bring them to the basis where the charged lepton Yukawa
coupling is diagonal will modify the values of the
soft terms calculated above. Defining
$\Ye=V_R{\Y}^{\rm diag}_e V_L^\dagger $, where ${\Y}^{\rm diag}_e$
is a diagonal real matrix and $V_{L,R}$ are unitary matrices,
the basis transformation 
$L \rightarrow V_L L$, $e_R \rightarrow V_R e_R$ yields
\begin{eqnarray}
\Ye &\rightarrow& {\Y}^{\rm diag}_e\;, \nonumber \\
\mes &\rightarrow& V_R^\dagger \mes V_R\;, \nonumber \\
\mLs &\rightarrow& V_L^\dagger \mLs V_L\;, \nonumber \\
\Ae &\rightarrow& V_R^\dagger \Ae V_L\;.
\label{diagonalization}
\end{eqnarray}

The explicit expression for $V_L$ is
\begin{eqnarray}
 V_L &\simeq & \left(\begin{array}{ccc}
 1           &  V_{L,12}  &  V_{L,13}\\
 V_{L,21}    &     1      &  V_{L,23}\\
-V_{L,13}^*  & -V_{L,23}^* &       1
\end{array} \right)\;,
\end{eqnarray}
where
\begin{eqnarray}
 V_{L,12} &=& \left( \fr \right)^2  \left[ \frac{3}{2} \lg^2 +2\lg \right] \YYnb{31}^* \YYnb{32}+{\cal O}(Y^6_{\nu})\;,\nonumber \\
 V_{L,21} &=& -\left( \fr \right)^2  \left[ \frac{5}{2} \lg^2 +2\lg \right] \YYnb{32}^* \YYnb{31}+{\cal O}(Y^6_{\nu})\;,\nonumber\\
 V_{L,13,23} &=&-\fr \lg \YYnb{31,32}^*+{\cal O}(Y^4_{\nu})\;.
\end{eqnarray}
Notice that $V_{L,21}\neq -V^*_{L,12}$, as required by unitarity. 
On the other hand, the expression for $V_R$ is
\begin{eqnarray}
 V_R &\simeq& \left(\begin{array}{ccc}
 1             &  V_{R,12}    & V_{R,13}\\
V_{R,21}    &    1         & V_{R,23}\\
-V_{R,13}^*    &  -V_{R,23}^* & 1
\end{array} \right)\;,
\end{eqnarray}
where
\begin{eqnarray} 
 V_{R,12} &=& \left(\frac{1}{8\pi^2}\right)^2 \frac{y_e}{y_\mu} 
\left[\left(1-\frac{y^2_\mu}{y^2_\tau}\right) \lg^2+  \lg \right] 
\YYnb{31}^* \YYnb{32}+{\cal O}(Y^6_{\nu})\;,\nonumber \\
 V_{R,21} &=& -\left(\frac{1}{8\pi^2}\right)^2 \frac{y_e}{y_\mu} 
\left[ \lg^2+  \lg \right] 
 \YYnb{32}^*\YYnb{31}+{\cal O}(Y^6_{\nu})\;,\nonumber \\
 V_{R,13} &=& -\frac{1}{8\pi^2}\frac{y_e}{y_\tau}
\lg \YYnb{31}^*+{\cal O}(Y^4_{\nu})\;,
\nonumber \\
 V_{R,23} &=& -\frac{1}{8\pi^2}\frac{y_\mu}{y_\tau}\lg 
\YYnb{32}^*+{\cal O}(Y^4_{\nu})\;.
\end{eqnarray}

With these expressions for $V_L$ and $V_R$ it is straightforward
to compute, using Eq.~(\ref{diagonalization}), the off-diagonal 
elements of the leptonic soft mass terms in the basis where the 
charged lepton Yukawa coupling is diagonal. 
In this ``mass basis'' the off-diagonal elements of the right-handed 
slepton mass matrix read:
\begin{equation}
(\mes)^{\rm mb}_{31,32}={\cal O}(Y^4_{\nu})\;,
~~~~~~(\mes)^{\rm mb}_{21}={\cal O}(Y^6_{\nu})\;,
\end{equation}
that no longer vanish, but still give negligible contributions
to the rare decays compared to the other sources of flavour violation.
On the other hand, for the left-handed slepton mass matrix they 
approximately read
\begin{eqnarray}
(\mLs)^{\rm mb}_{31,32}(\mmaj)&\simeq& -\frac{1}{8\pi^2}
(m_L^2+m_\nu^2+\mHus+|\Ann|^2) \YYnb{31,32}\lg\;,
\nonumber\\
(\mLs)^{\rm mb}_{21}(\mmaj) &\simeq& \left(\fr\right)^2 
\Bigg[  8 (m_L^2+\mHus+m_\nu^2+2 |\Ann|^2) \lg  +\nonumber\\
&&\hspace{-2cm}8( m_L^2+ \mHus+ m_\nu^2+\frac{5}{2} |\Ann|^2)\lg^2 \Bigg]
\YYnb{32}^* \YYnb{31} \;.\hspace{1cm}
\label{final-scalar}
\end{eqnarray}
Note that the diagonalization of the charged-lepton Yukawa coupling
does not alter $\mLs(\mmaj)_{31,32}$ but only $(\mLs)(\mmaj)_{21}$.
Finally,  the off-diagonal elements of the
soft trilinear term read:
\begin{eqnarray}
(\Ae)^{\rm mb}_{31,32} &\simeq& - \fr \lg 2\Ann y_\tau \YYnb{31,32}\;,
\nonumber \\
(\Ae)^{\rm mb}_{13,23} &\simeq& - \fr \lg 2\Ann h_{e,\mu} \YYnb{31,32}^* 
\;,\nonumber \\
 (\Ae)^{\rm mb}_{21} &\simeq& \left(\fr \right)^2 
  \Bigg[8\lg+ 8\lg^2 \Bigg]\Ann y_{\mu} \YYnb{32}^* \YYnb{31} \;,
\nonumber \\
 (\Ae)^{\rm mb}_{12} &\simeq& \left(\fr \right)^2 
  \Bigg[8\lg+ 8\lg^2 \Bigg]\Ann y_e \YYnb{31}^* \YYnb{32}\;. \hspace{1cm}
\label{final-trilinear}
\end{eqnarray}
In this case, {\it all} the off-diagonal elements get modified
at the lowest order by the basis transformation.

It is apparent from Eqs.~(\ref{final-scalar},\ref{final-trilinear}) 
that in the scenario with flavour universality of the soft terms at $\mgut$ 
and $\YYnb{21}(\mgut)=0$ there exists a very precise correlation between 
the 21 and the 31 and 32 entries of the soft terms. As was
argued in the previous section, this scenario yields the minimal amount
of flavour violation in the $\mu-e$ sector. Therefore, any
other neutrino Yukawa coupling compatible with the experimental data 
will induce larger $(\mLs)^{\rm mb}_{21}$ and $(\Ae)^{\rm mb}_{12,21}$ entries
at low energies. For example, under the assumption of complete
universality of the soft mass terms at $\mgut$, 
$m_L^2(\mgut)=\mHus(\mgut)=m_\nu^2(\mgut)\equiv m^2_0$,
the following lower bounds hold at low energies for the 21 entries
of the soft terms:
\begin{eqnarray}
\left|\frac{(\mLs)^{\rm mb}_{21}}{
 (\mLs)^{\rm mb}_{31}(\mLs)^{\rm mb}_{32}}\right|
&\gsim& \frac{2(3m^2_0+2|\Ann|^2)\lg
+2(3m^2_0+\frac{5}{2}|\Ann|^2)\lg^2}{\left[(3m^2_0+|\Ann|^2)\lg\right]^2}\;,
\nonumber \\
\left|\frac{(\Ae)^{\rm mb}_{21}}{(\Ae)^{\rm mb}_{23}(\Ae)^{\rm mb}_{31}}\right|
&\simeq&
\left| \frac{(\Ae)^{\rm mb}_{12}}{(\Ae)^{\rm mb}_{13}(\Ae)^{\rm mb}_{32}}\right|
\gsim \frac{2}{ y_\tau|\Ann|}\left(1+\frac{1}{\lg}\right)\;.
\label{soft-bounds}
\end{eqnarray}

These bounds will eventually translate into a bound 
on $BR(\mu\rightarrow e\gamma)$ involving the branching
ratios of the rare tau decays.

\section{Lower bound on $\mu\rightarrow e\gamma$ from 
rare tau decays}
After computing the radiatively generated off-diagonal elements of
the soft SUSY breaking terms, it is straightforward to calculate
the branching ratios for the rare lepton decays. In order to
understand qualitatively the results we will use in this section
approximate formulas for the branching ratios, although in our numerical
analysis we used the general expressions existing in the 
literature~\cite{Hisano:1995cp} and we solved numerically 
the renormalization group equations including the two-loop 
$\beta$-functions.

A very useful tool to treat analytically the complicated exact
expressions for the branching ratios is the mass insertion approximation,
where the small off-diagonal elements of the soft terms are treated as
insertions in the sfermion propagators in the loops~\cite{FCNC,Hisano:1998fj}. 
This rationale can also be applied to the gaugino-higgsino sector, 
yielding at the end of the day relatively compact expressions for 
the branching ratios. 

It was argued in~\cite{Hisano:1998fj} that for the rare 
tau decays, the dominant contributions correspond to the mass 
insertion diagrams enhanced by $\tan\beta$ factors. Namely, 
\begin{eqnarray}
BR(\tau\rightarrow e\gamma) &\simeq& \frac{\alpha^3}{G_F^2}
\frac{|(\mLs)_{31}|^2}{m_S^8}\tan^2\beta 
BR(\tau\rightarrow e \nu_\tau \bar \nu_e) \nonumber \\
BR(\tau\rightarrow \mu\gamma) &\simeq& \frac{\alpha^3}{G_F^2}
\frac{|(\mLs)_{32}|^2}{m_S^8}\tan^2\beta 
BR(\tau\rightarrow \mu \nu_\tau \bar \nu_\mu) 
\label{BRtaugamma}
\end{eqnarray}
where $BR(\tau\rightarrow \mu \nu_\tau \bar \nu_\mu)\simeq 0.17$,
$BR(\tau\rightarrow e \nu_\tau \bar \nu_e)\simeq 0.18$ and
$m_S$ is a mass scale of the order of typical SUSY masses.
In the case of the
Constrained MSSM it is best approximated by $m_S^8\simeq 0.5 
m_0^2 M^2_{1/2}(m_0^2+0.6 M^2_{1/2})^2$~\cite{Petcov:2003zb},
where $m_0$ is the universal scalar mass and $M_{1/2}$ is
the universal gaugino mass at $\mgut$.

On the other hand, being $|(\mLs)_{21}|$ and $|(\Ae)_{21}|$ 
generated at order ${\cal O}(Y^4_{\nu})$, we keep for consistency 
the contributions to $BR(\mu\rightarrow e\gamma)$ induced
not only by a single mass insertion 
(where $(\mLs)_{21}$ or $(\Ae)_{21}$ is inserted)
but also by a double mass insertion (where $(\mLs)_{32}$ or $(\Ae)_{32}$, 
and $(\mLs)_{31}$ or $(\Ae)_{13}$ are inserted). The result
reads:
\begin{equation}
BR(\mu\rightarrow e\gamma) \simeq \frac{\alpha^3}{G_F^2}
\left|
\frac{(\mLs)_{21}}{m_S^4}-
\frac{(\mLs)_{32}^*(\mLs)_{31}}{m_{S^\prime}^{6}}\right|^2\tan^2\beta \;,
\label{BRmuegamma}
\end{equation}
where $m_{S^\prime}$ is another mass scale of the order of $m_{S}$, 
although in general different (note that the single and the double mass 
insertion contributions have opposite signs). 
Inserting the bound Eq.~(\ref{soft-bounds}) 
into Eq.~(\ref{BRmuegamma}), we obtain the following bound:
\begin{equation}
BR(\mu\rightarrow e\gamma) \gsim \frac{\alpha^3}{G_F^2}
\frac{|(\mLs)_{32}|^2|(\mLs)_{31}|^2}{m_{S^{\prime\prime}}^{12}} 
\tan^2\beta  \;,
\label{boundmeg}
\end{equation}
where $m_{S^{\prime\prime}}$ is another mass scale, again of the same
order of $m_{S}$, $m_{S^{\prime}}$. Using the expressions for
the rare tau decays in terms of $|(\mLs)_{31}|^2$, $|(\mLs)_{32}|^2$,
Eq.~(\ref{BRtaugamma}), this bound can be casted as:
\begin{equation}
BR(\mu\rightarrow e\gamma) \gsim \frac{G_F^2}{\alpha^3 \tan^2\beta}
\frac{m^{16}_S}{m^{12}_{S^{\prime\prime}}}\frac{BR(\tau\rightarrow \mu\gamma)}
{BR(\tau\rightarrow \mu \nu_\tau \bar \nu_\mu)}
\frac{BR(\tau\rightarrow e\gamma)}
{BR(\tau\rightarrow e \nu_\tau \bar \nu_e)} \;.
\label{boundmeg2}
\end{equation}
This equation is a more explicit expression of Eq.~(\ref{mainresult})
and constitutes the main result of this paper\footnote{One loop QED
corrections to the electric and magnetic dipole operators reduce
the theoretical prediction for $BR(l_j\rightarrow l_i \gamma)$ by a 
factor $\left(1-\frac{8\alpha}{\pi}
\log\frac{m_S}{m_j}\right)$~\cite{Czarnecki:2001vf}.
This correction makes the bound Eq.~(\ref{boundmeg2}) a 2-6\% stronger
for $m_S=100-1000$ GeV.}.

The numerical values of $m_S$ and $m_{S^{\prime\prime}}$ depend
crucially on the supersymmetric scenario. Before presenting
the exact numerical results for some common supersymmetric
scenarios, let us obtain first a rough estimate of the 
bound Eq.~(\ref{boundmeg2}). To this end, we will make
the approximation $m_{S^{\prime\prime}}\sim m_S$. Then, 
the previous bound reads:
\begin{equation}
BR(\mu\rightarrow e\gamma) \gsim  10^{-9}
\left(\frac{m_S}{200\,{\rm GeV}}\right)^4
\left(\frac{\tan\beta}{10}\right)^{-2}
\left(\frac{BR(\tau\rightarrow \mu\gamma)}{6.8\times10^{-8}}\right)
\left(\frac{BR(\tau\rightarrow e\gamma)}{1.1\times10^{-7}}\right)\;.
\label{boundmeg-rough}
\end{equation}
Therefore, if the rates for the rare tau decays were just below
the present experimental bounds, the see-saw scenario with 
universal soft terms at $\mgut$ would predict a branching ratio for 
$\mu\rightarrow e\gamma$ typically larger than $10^{-9}$,
in conflict with the present bound
$BR(\mu\rightarrow e\gamma)\leq 1.2\times 10^{-11}$. 
Furthermore, if the observation of {\it both} rare tau decays is
at the reach of present $B$-factories, $BR(\tau\rightarrow \l_i \gamma)$
have to be larger than $\sim 10^{-8}$, which would imply 
$BR(\mu\rightarrow e\gamma)\gsim 2\times 10^{-11}$, barely
compatible with the present upper bound from MEGA. One can make this
argument more quantitative turning Eq.~(\ref{boundmeg-rough}) 
to obtain an upper bound on $BR(\tau\rightarrow e\gamma)$, 
or alternatively on $BR(\tau\rightarrow \mu\gamma)$, from the stringent 
experimental constraint on $BR(\mu\rightarrow e\gamma)$:
\begin{eqnarray}
BR(\tau\rightarrow e\gamma)\hspace{-0.3cm}&\lsim&\hspace{-0.3cm} 10^{-9}
\left(\frac{m_S}{200\,{\rm GeV}}\right)^{-4}
\left(\frac{\tan\beta}{10}\right)^{2}
\left(\frac{BR(\mu\rightarrow e\gamma)}{1.2\times 10^{-11}}\right)
\left(\frac{BR(\tau\rightarrow \mu\gamma)}{6.8\times 10^{-8}}\right)^{-1}
\hspace{-0.4cm}\;,
\nonumber\\
BR(\tau\rightarrow \mu\gamma)\hspace{-0.3cm}&\lsim& \hspace{-0.3cm}7\times 10^{-10}
\left(\frac{m_S}{200\,{\rm GeV}}\right)^{-4}
\left(\frac{\tan\beta}{10}\right)^{2}
\left(\frac{BR(\mu\rightarrow e\gamma)}{1.2\times 10^{-11}}\right)
\left(\frac{BR(\tau\rightarrow e\gamma)}{1.1\times 10^{-7}}\right)^{-1}
\hspace{-0.4cm}\;.\hspace{0.9cm}
\end{eqnarray}
Then, if $BR(\tau\rightarrow \mu\gamma)\gsim 10^{-8}$, 
thus making the observation
of $\tau\rightarrow \mu\gamma$ accessible to present $B$-factories,
the previous equation would imply that 
$BR(\tau\rightarrow e\gamma)\lsim 8\times 10^{-9}$,
which would make the observation of $\tau\rightarrow e\gamma$
difficult (analogous conclusions can be drawn 
for $BR(\tau\rightarrow \mu\gamma)$).
Therefore, whereas the observation of one rare tau decay 
at present $B$-factories is indeed possible, in the see-saw 
scenario the observation of {\it both} rare tau decays is unlikely. 

This qualitative discussion shows that in the supersymmetric 
see-saw model the so far negative searches 
for $\mu\rightarrow e\gamma$ have crucial implications for the
future searches for $\tau\rightarrow \mu\gamma$
and $\tau\rightarrow e\gamma$. Clearly, these implications will become 
stronger as the MEG experiment at PSI improves the bound on 
$\mu\rightarrow e\gamma$. However, one should bear in mind that 
this is just a qualitative 
discussion and that the actual impact of the bound Eq.~(\ref{boundmeg2})
on future searches for rare decays depends on the particular 
point of the SUSY parameter space, through the values of $m_S$ 
and $m_{S^{\prime\prime}}$. We will see,
however, that these strong conclusions hold for a wide choice 
of supersymmetric parameters. 

We have investigated in detail the 'Snowmass Points and Slopes' 
(SPS)~\cite{Allanach:2002nj}, which constitute a set of benchmark 
points in the supersymmetric parameter space aiming to describe typical
points, but also extreme although well motivated possibilities.
We have analyzed the six mSUGRA benchmark points, that are defined
at $\mgut=2\times 10^{16}$ GeV by five parameters: the
universal scalar mass ($m_0$), gaugino mass ($M_{1/2}$) and
trilinear term ($A_0$), $\tan\beta$ and the sign of $\mu$.
For each SPS point, the values of these parameters are given in
Table 2.

\begin{table}[h]
\begin{center}
\begin{tabular}{|c|ccccc|}
\hline 
& $m_0$  (GeV)& $M_{1/2}$ (GeV) & $A_0$ (GeV) & $\tan\beta$ & ${\rm sign}\;\mu$ \\ \hline 
SPS1a &  100 & 250 & -100 & 10 & +  \\
SPS1b &  200 & 400 &    0 & 30 & +  \\
SPS2  & 1450 & 300 &    0 & 10 & +  \\
SPS3  &   90 & 400 &    0 & 10 & +  \\
SPS4  &  400 & 300 &    0 & 50 & +  \\
SPS5  &  150 & 300 &-1000 &  5 & +  \\ \hline 
\end{tabular}
\end{center}
\caption{\small Parameters at $\mgut$ for the six 
mSUGRA SPS benchmark points.} 
\end{table}

The SPS1a and SPS1b points are ``typical'' mSUGRA points with 
intermediate and relatively high values of $\tan\beta$
respectively. Assuming that the neutralino constitutes
the dominant component of dark matter of the Universe,
these two points lie in the ``bulk'' region of the allowed
mSUGRA parameter space. On the other hand, 
the SPS2 point is characterized by heavy squarks and sleptons and
fairly light neutralinos, charginos and gluinos\footnote{The 
low energy predictions for this benchmark point are
extremely sensitive to the value of the top quark mass~\cite{Allanach:2003jw}.
In particular, assuming $M_t=175$ GeV this point
would lie in the ``focus point'' region of the allowed mSUGRA 
parameter space. However, the most recent measurement
of the top quark mass at CDF II, $M_t=170.8\pm2.2 ({\rm stat.})
\pm1.4({\rm syst.})$ GeV~\cite{Abulencia:2007br},
pushes the SPS2 benchmark point out of the ``focus point'' region.}.
The SPS3 point has a small stau-neutralino mass difference
and lies in the stau coannihilation region. The SPS4 point
has a large $\tan\beta$ and lies in the ``funnel'' region.
Lastly, the SPS5 point is characterized by a relatively 
light stop.
We have also analyzed for completeness the mSUGRA-like scenario
SPS6, characterized by having non-universal gaugino masses,
and defined at the GUT scale by $m_0=150$ GeV, $M_3=M_2=300$ GeV,
$M_1=480$ GeV, $A_0=0$, $\tan\beta=10$ and positive $\mu$.

In Fig.~\ref{figure1} we show the allowed values for 
$BR(\tau\rightarrow e \gamma)$ and $BR(\tau\rightarrow \mu \gamma)$
in the supersymmetric see-saw model for the mSUGRA benchmark 
points SPS1-5. The area above (to the right of) the dashed line
at $BR(\tau\rightarrow \mu \gamma)=6.8\times 10^{-8}$ 
($BR(\tau\rightarrow e \gamma)=1.1\times 10^{-7}$) is excluded by the present
experimental bounds on the rare tau decays.
On the other hand, the area above the diagonal line labeled 
$BR(\mu\rightarrow e\gamma)<1.2 \times 10^{-11}$ 
is excluded from the present experimental bound
on $\mu\rightarrow e\gamma$, as a consequence of
Eq.~(\ref{boundmeg2}). 
We find remarkable that for {\it all} the mSUGRA SPS points 
the bound Eq.~(\ref{boundmeg2}) excludes
values for the branching ratios of the rare
tau decays that are otherwise allowed by 
direct searches. Furthermore, if the MEG experiment
reaches the sensitivity $BR(\mu\rightarrow e \gamma)\sim 10^{-13}$
without observing a positive signal, the region of the parameter
space excluded by Eq.~(\ref{boundmeg2}) would enlarge considerably.
Finally, in the plots we assumed an intermediate decoupling scale,
$\mmaj=5\times10^{13}$ GeV. Had we used a larger value for
$\mmaj$, the excluded region would also enlarge, as is apparent
from Eq.~(\ref{soft-bounds})\footnote{Recall that to bring the decay rates
to the reach of present and future experiments the neutrino
Yukawa couplings have to be sizable. This typically requires large values
of $\mmaj$ in order to produce small neutrino masses.}.

The bound Eq.~(\ref{boundmeg2}) also
has implications for future searches for rare
tau decays. In Fig.~{\ref{figure1} we show with
a dash-dotted line the projected sensitivity of present
$B$-factories to rare tau decays ($BR(\tau\rightarrow \mu \gamma),
BR(\tau\rightarrow e \gamma)\gsim 10^{-8}$). Therefore,
the area shaded in green is the region of this parameter
space accessible to present $B$-factories, that we call
for definiteness the ``observable window of present $B$-factories''.
Using Eq.~(\ref{boundmeg2}) we find that large regions of the 
observable window of present $B$-factories are excluded from
the present bound $BR(\mu\rightarrow e \gamma)<1.2\times 10^{-11}$
(light green shaded area).
In particular, for the benchmark points SPS1a, SPS1b and SPS3 
the region where {\it both} $\tau\rightarrow \mu \gamma$ 
and $\tau\rightarrow e \gamma$ could be discovered at present
$B$-factories is excluded. Nevertheless, the discovery of
one of them, either $\tau\rightarrow \mu \gamma$  or 
$\tau\rightarrow e \gamma$, is still possible.
Let us remind that SPS1a and SPS1b
correspond to ``typical'' mSUGRA points, and thus this conclusion
holds for a large region of the parameter space. 
On the other hand, for SPS2 and SPS5
the discovery of both rare decays is still possible, although
only if their branching ratios are close to the experimental
sensitivity of present $B$-factories\footnote{One should bear 
in mind, however, that the SPS2 
point is extremely sensitive to the input value of the top quark 
mass. Therefore, the conclusions for this benchmark point 
should be taken with a pinch of salt.}. Finally, for SPS4 the present 
bound $BR(\mu\rightarrow e \gamma)<1.2\times 10^{-11}$ 
has only little impact for present $B$-factories. 
An improvement of the bound on $BR(\mu\rightarrow e \gamma)$
by two orders of magnitude, as planned by the MEG experiment at PSI, 
would exclude the possibility of observing both rare tau decays 
at present $B$-factories in most mSUGRA parameter space.
Hence, would present $B$-factories refute this expectation,
the supersymmetric see-saw model with mSUGRA
would have to be abandoned, unless a certain 
amount of fine tuning is accepted. The same conclusions
hold for the mSUGRA-like scenario SPS6, which has non-universal
gaugino masses, as can be realized from Fig.~\ref{figure2}.

We also show for completeness the observable window of the projected
super$B$-factories, shown as a yellow shaded area. It is defined
by the region between the dashed lines indicating the present 
bounds on the rare tau decays and the dotted 
lines showing the projected bounds ($BR(\tau\rightarrow \mu \gamma),
BR(\tau\rightarrow \mu \gamma)\gsim 10^{-9}$). The present bound
on $\mu\rightarrow e\gamma$ practically does not exclude
any region of the observable window of the projected
super$B$-factories, except for the benchmark point SPS3.
On the other hand, if the MEG experiment reaches the projected 
sensitivity $BR(\mu\rightarrow e \gamma)\sim 10^{-13}$
without observing a positive signal, again a large portion
of the observable window of the projected super$B$-factories
would be excluded by Eq.~(\ref{boundmeg2}). In particular, 
whereas the observation of either $\tau\rightarrow \mu \gamma$
or $\tau\rightarrow e \gamma$ will indeed be possible at 
the projected super$B$-factories, the observation of both would
only be possible for the benchmark point SPS4 and marginally
for SPS2 and SPS5. For the rest of the benchmark points analyzed in this
paper, SPS1a, SPS1b, SPS3 and SPS6, the observation
of both rare decays will not be possible in the supersymmetric see-saw
model, unless a certain amount of fine tuning is accepted.

\begin{figure}
\begin{center}
\begin{tabular}{c}
\epsfig{figure=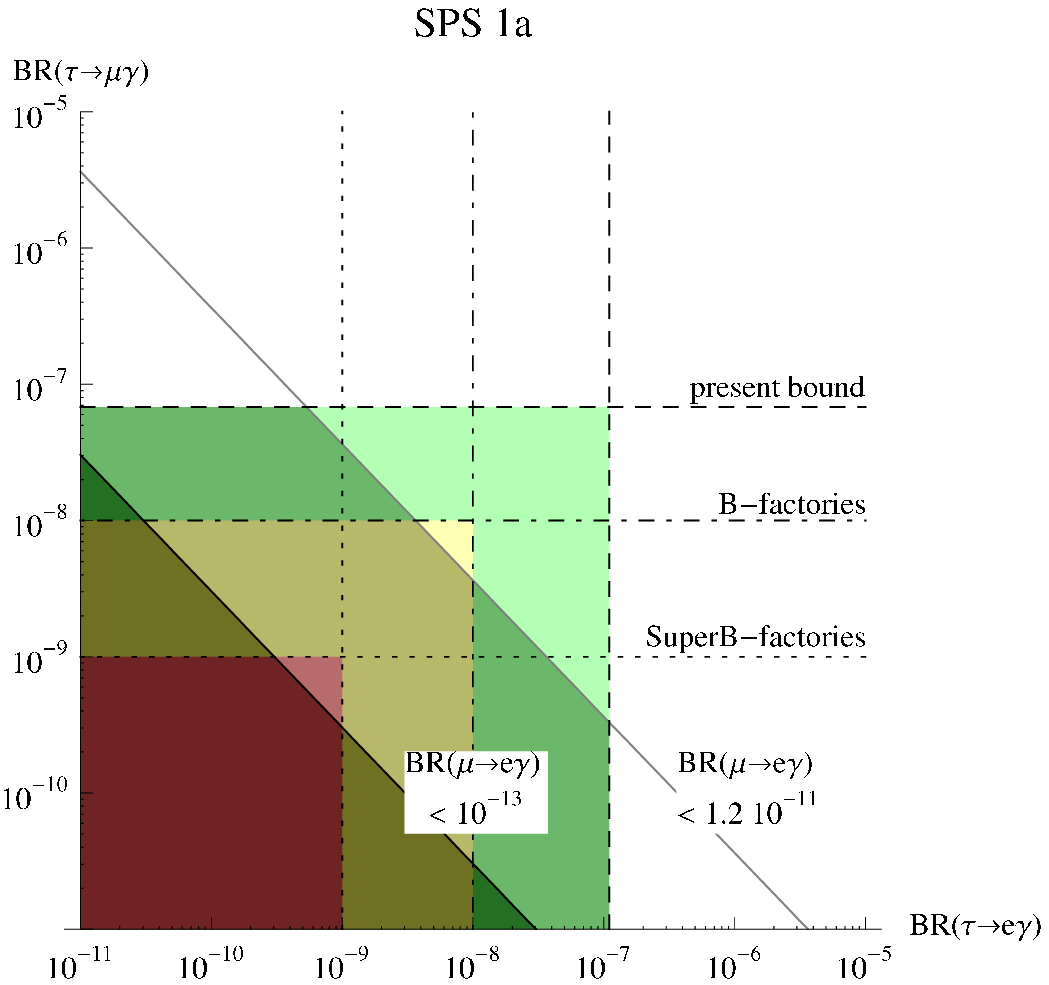,width=65mm} 
\epsfig{figure=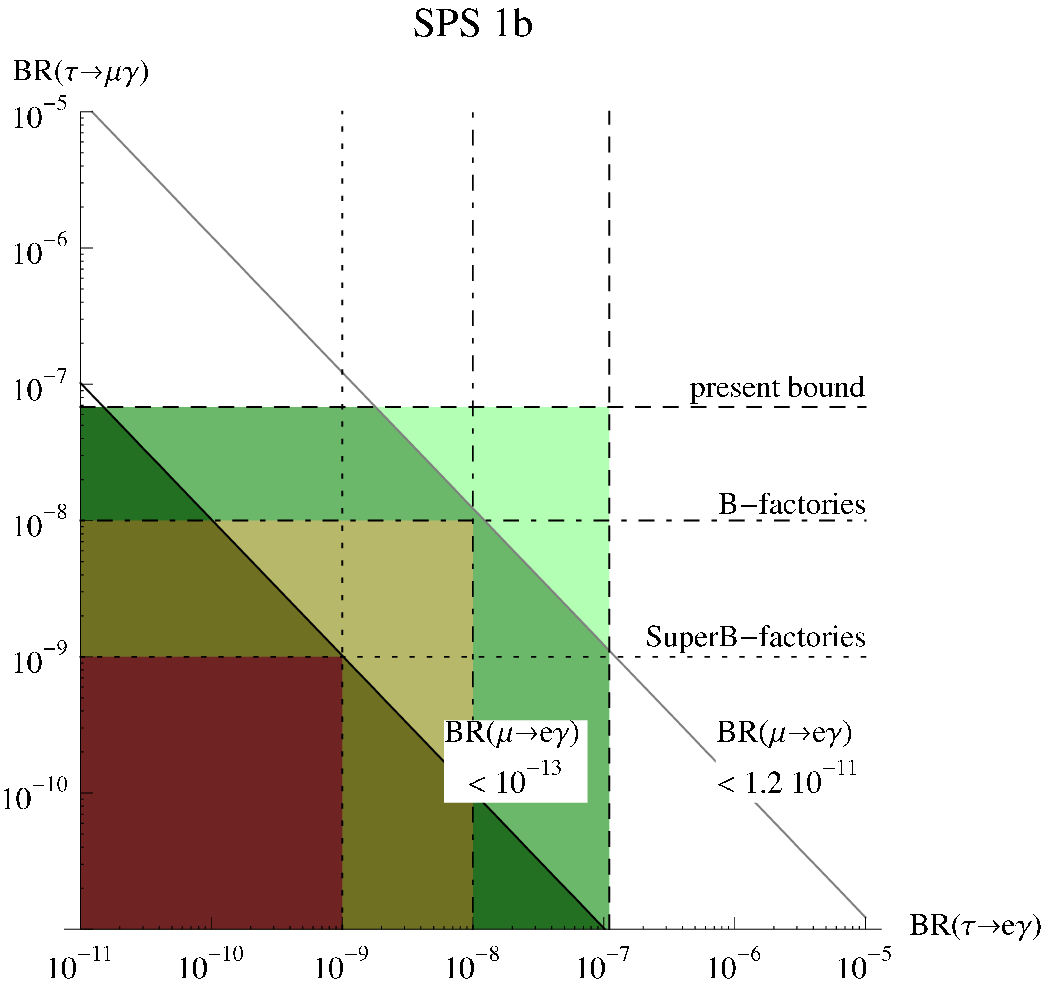,width=65mm}\\\\
\epsfig{figure=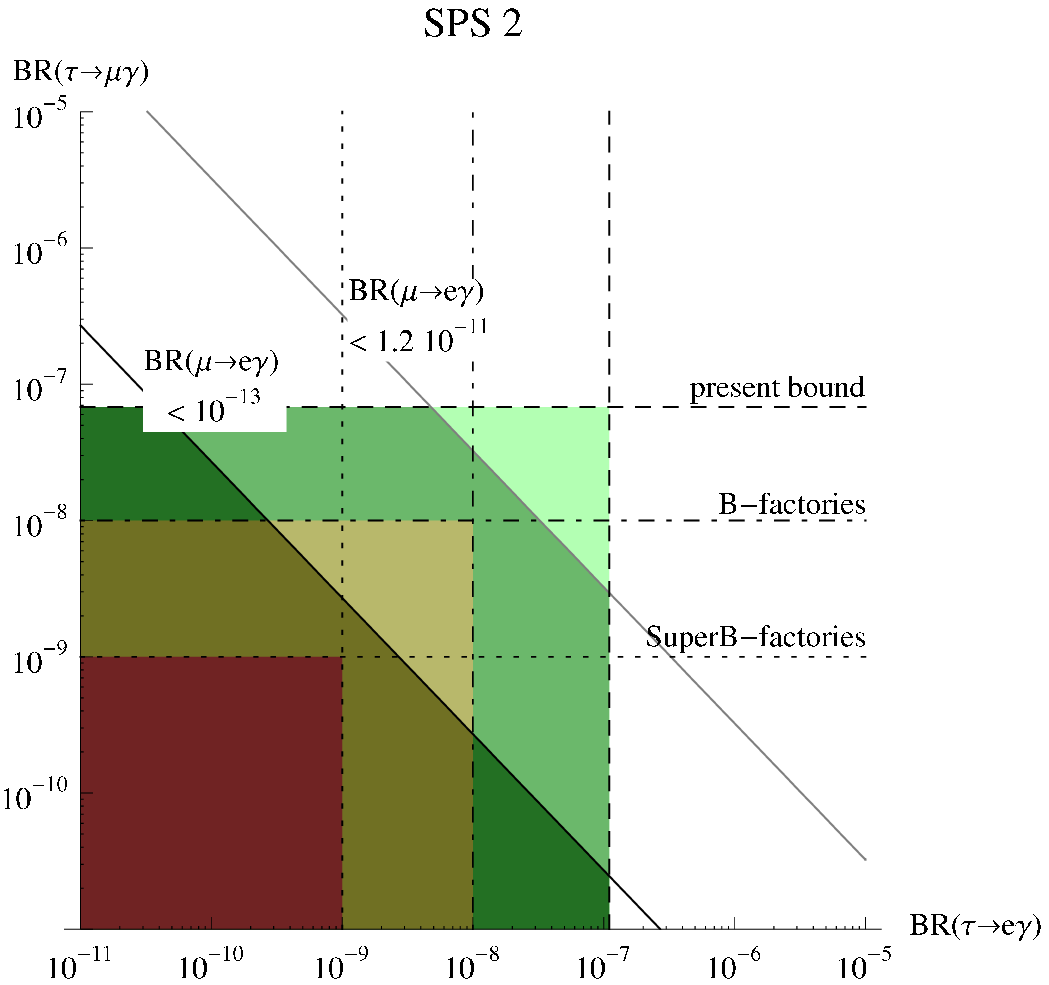,width=65mm} 
\epsfig{figure=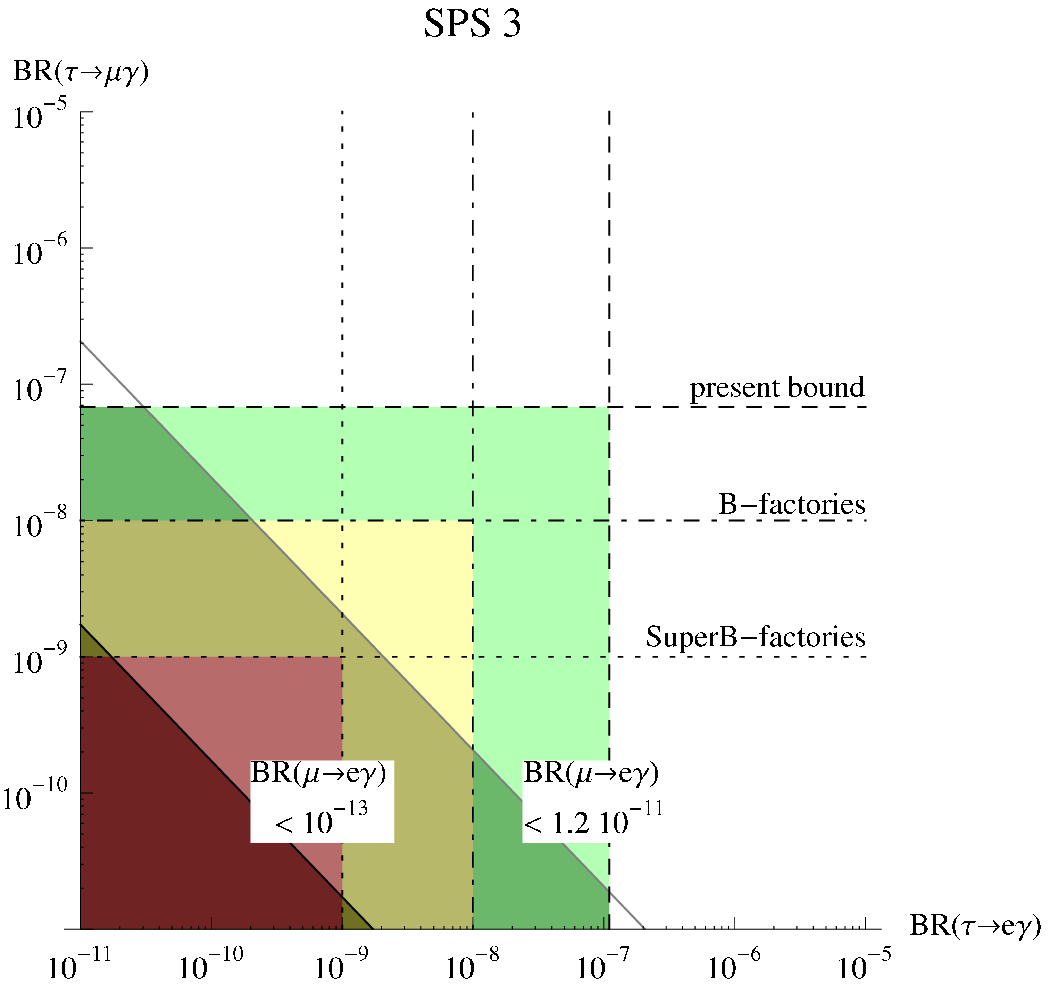,width=65mm} \\\\
\epsfig{figure=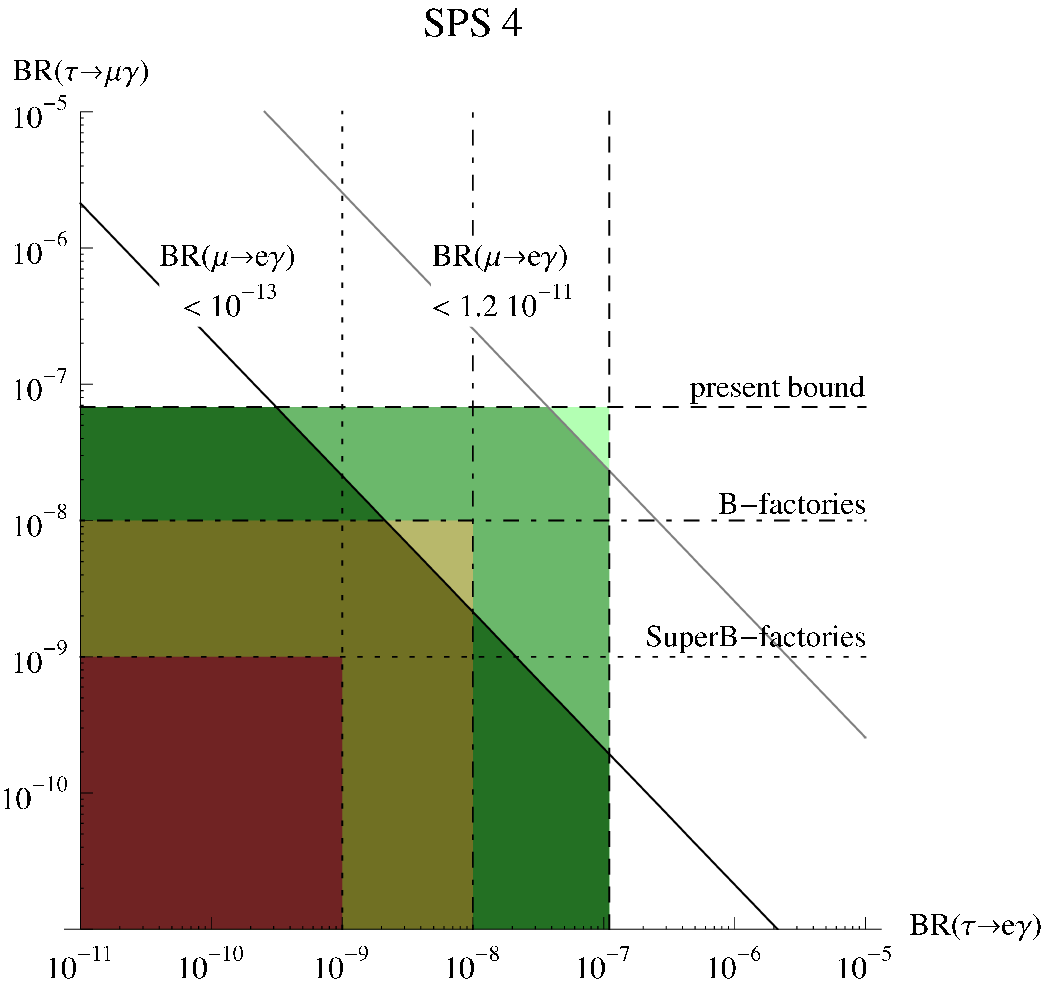,width=65mm} 
\epsfig{figure=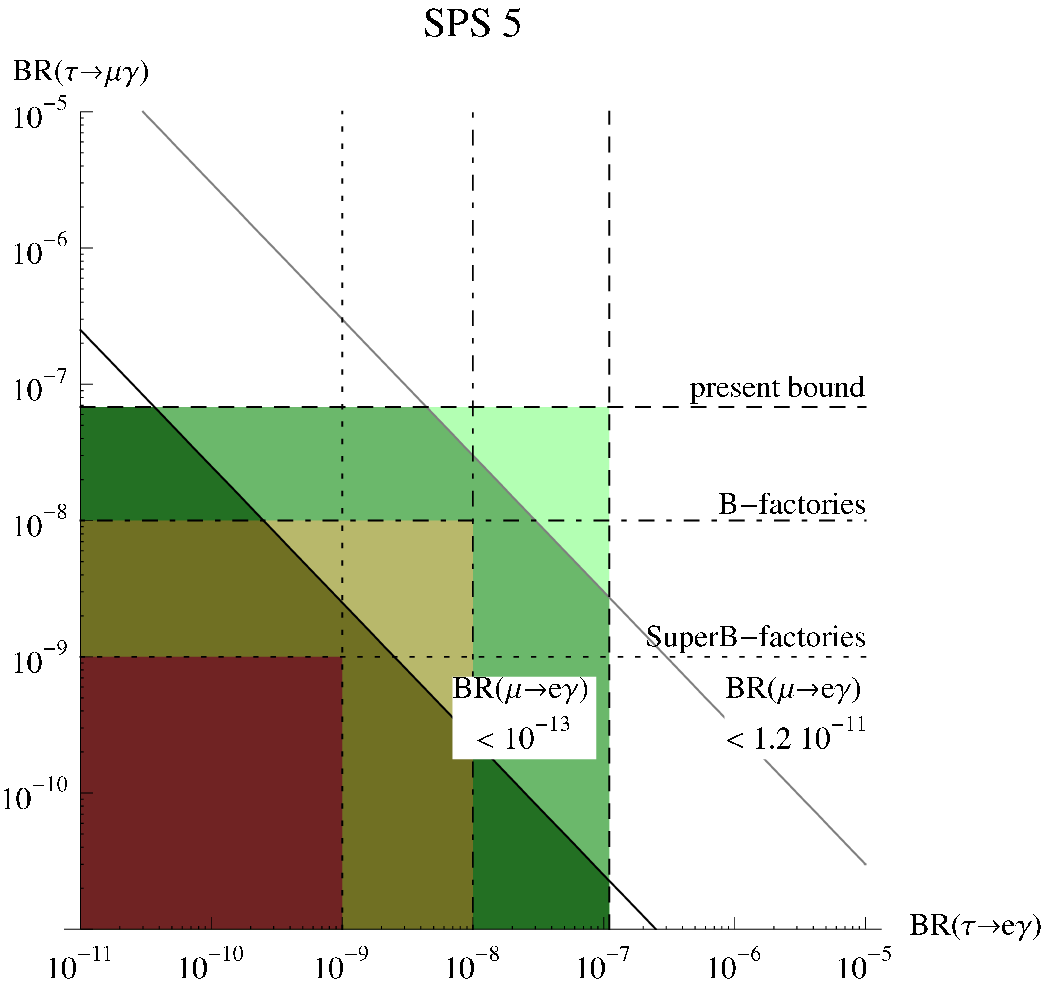,width=65mm} 
\end{tabular}
\end{center}
\caption
{\small Allowed values for the branching ratios
of the rare tau decays $\tau\rightarrow e\gamma$ and
$\tau\rightarrow \mu\gamma$ from present experiments
and from the bound $BR(\mu\rightarrow e\gamma)\gsim
C \times BR(\tau\rightarrow \mu\gamma)BR(\tau\rightarrow e\gamma)$
for the mSUGRA scenarios SPS1-5. The area in green
indicates the observable window of present $B$-factories,
and in yellow the observable window of future super$B$-factories.
Excluded regions are shown with light shading, whereas
allowed regions are shown with dark shading. In the plots
it is assumed $\mmaj=5\times 10^{13}$ GeV.
}
\label{figure1}
\end{figure}

\begin{figure}
\begin{center}
\begin{tabular}{c}
\epsfig{figure=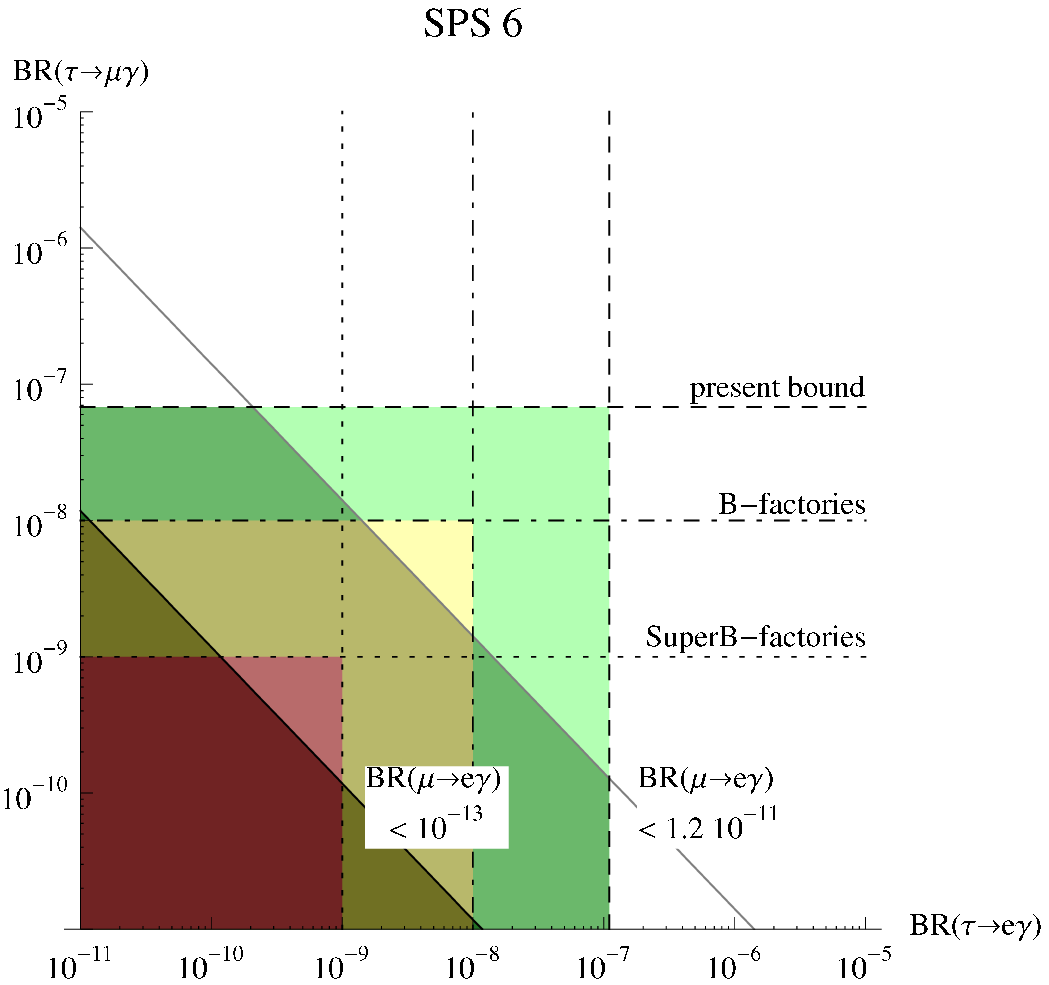,width=65mm}
\end{tabular}
\end{center}
\caption
{\small As in Fig.~\ref{figure1}, but for the mSUGRA-like scenario SPS6.}
\label{figure2}
\end{figure}

\section{Conclusions}

A series of neutrino experiments have demonstrated that 
{\it all} family lepton numbers are violated in Nature. 
Therefore, there is no symmetry reason forbidding 
the rare lepton decays $\mu\rightarrow e\gamma$, 
$\tau\rightarrow \mu \gamma$ or $\tau\rightarrow e\gamma$.
In this paper we have discussed the implications of
this observation for the supersymmetric see-saw model.

We have shown that even in the very special situation where the
flavour violation vanishes at high-energies in one of the sectors, 
for instance in the $\mu-e$ sector, the flavour violation
in the other two sectors will induce through two loop radiative
corrections a small though non-vanishing
flavour violation at low energies in the $\mu-e$ sector,
that will be correlated to the flavour violation in the $\tau-\mu$
and $\tau-e$ sectors.
Barring cancellations, this scenario will produce the
minimal rate  for the rare decay $\mu\rightarrow e\gamma$.
Therefore, in any other scenario the lower bound
$BR(\mu\rightarrow e\gamma)\gsim
C \times BR(\tau\rightarrow \mu\gamma)BR(\tau\rightarrow e\gamma)$
will hold, where $C$ is a constant that depends on supersymmetric 
parameters.

We have analyzed the implications of this bound on the possible
values of the rates of the rare tau decays for the
supersymmetric benchmark points SPS1-6. We have found that
values for $BR(\tau\rightarrow \mu\gamma)$ and 
$BR(\tau\rightarrow e\gamma)$ that are allowed by present
experiments searching for rare tau decays are forbidden
by our bound $BR(\mu\rightarrow e\gamma)\gsim
C \times BR(\tau\rightarrow \mu\gamma)BR(\tau\rightarrow e\gamma)$
and the present constraint on $BR(\mu\rightarrow e\gamma)$ from MEGA.
In particular, we have found that, for large regions of the
Constrained MSSM parameter space, present $B$-factories
could discover either $\tau\rightarrow \mu\gamma$ or 
$\tau\rightarrow e\gamma$, but not both. 
We have also discussed the implications of the non-observation
of the process $\mu\rightarrow e\gamma$ by the MEG experiment at PSI
for the future searches for
rare tau decays at present $B$-factories and at projected 
super$B$-factories. This analysis could also be extended to more
general classes of models. Work along these lines is 
already in progress~\cite{ISS}.

\section*{Acknowledgements}
We are grateful to Alberto Casas, Andrzej Czarnecki, 
Michael Ratz and especially to
Tetsuo Shindou for useful discussions and suggestions.
This research was supported by the DFG cluster of
excellence Origin and Structure of the Universe  and by the 
SFB-Transregio 27 "Neutrinos and Beyond".

\section*{Appendix}
In this appendix we report the full set of one- and two-loop 
Renormalization Group Equations (RGEs) for the parameters
of the Minimal Supersymmetric Standard Model extended with
three right-handed neutrino superfields. Partial results
for the Yukawa couplings, the gauge couplings and 
the neutrino mass matrix can be found in \cite{RGEs-partial}.
The RGEs for the soft SUSY breaking terms and the $\mu$ term
have been derived particularizing the general formulas in 
\cite{Martin:1993zk} to the supersymmetric see-saw 
model\footnote{All the RGEs are written in the $\overline{DR}^\prime$ 
scheme.}.

The RGE for any supersymmetric or soft-breaking parameter 
can be schematically written as
\begin{equation}
 \frac{d}{d t} X = \fr \beto_X + \frsq \bett_X\;.
\end{equation}
The one- and two-loop $\beta$ functions for the gauge couplings 
are given by
\begin{align}
\beto_{g_a}  &= g_a^3 B^{(1)}_a\;,\\
\bett_{g_a} &= g_a^3 \left [\sum_{b=1}^3 B^{(2)}_{ab} g_b^2 - \sum_{x=u,d,e,\nu} C_a^x \, \Tr (\Yx^\dag\Yx) \right ]\;,
\end{align}
where $g_1$, $g_2$ and $g_3$ are the $U(1)_{Y}$, $SU(2)_{L}$ and $SU(3)_C$ 
gauge couplings, respectively, and
\begin{align*}
B_a^{(1)} &= (33/5,1,-3)\;, \eqnum \label{defs-gauge}\\
B^{(2)}_{ab} &= \begin{pmatrix} {199/ 25} & {27/5} & {88/ 5} \\
                 {9/ 5}    & 25      & 24          \\
                 {11/ 5}   &  9          & 14      \end{pmatrix}\;,\\
C^{u,d,e,\nu}_a &= \begin{pmatrix}  {26/ 5} & {14/ 5} & {18/ 5}  & 6/5\\
                          6         & 6           & 2      & 2     \\
                          4         & 4           & 0      & 0 \end{pmatrix}\;.
\end{align*}
(In these expressions, $g_1$ has been normalized as in $SU(5)$.)

On the other hand, the complete superpotential of the MSSM 
extended with right-handed neutrinos reads, imposing R-parity conservation,
\begin{equation}
W= {d_R^c}_i \Yd_{ij} Q_j  H_d+
{u_R^c}_i \Yu_{ij} Q_j  H_u+
{e_R^c}_i \Ye_{ij} L_j  H_d+
{\nu_R^c}_i \Yn_{ij} L_j  H_u+\mu   H_u H_d
- \frac{1}{2}{\nu_R^c}_i{\sM}_{ij}{\nu_R^c}_j \;.
\end{equation}
The one- and two-loop $\beta$ functions for these
SUSY conserving parameters are:
\begin{align*}
\beto_{\Yd} =
\Yd \biggl\lbrace
&\Tr (3 \Yd \Yd^\dag +\Ye \Ye^\dag)+ 3 \Yd^\dag \Yd + \Yu^\dag \Yu
  - \frac{16}{ 3} g_3^2 - 3 g_2^2 - \frac{7}{ 15} g_1^2
\biggr \rbrace \;,
\eqnum\displaybreak[2]\\
\bett_{\Yd} = 
\Yd \biggl\lbrace
&-\Tr (9\Yd \Yd^\dag \Yd \Yd^\dag + 3\Yu \Yd^\dag \Yd \Yu^\dag
  + 3\Ye \Ye^\dag \Ye \Ye^\dag + \Yn \Ye^\dag \Ye \Yn^\dag)
\eqnum\\&- \Yu^\dag \Yu \Tr (3\Yu \Yu^\dag + \Yn \Yn^\dag)
  - 3 \Yd^\dag \Yd \Tr (3 \Yd \Yd^\dag + \Ye \Ye^\dag)
\\&- 4 \Yd^\dag \Yd \Yd^\dag \Yd - 2 \Yu^\dag \Yu \Yu^\dag \Yu
  - 2 \Yu^\dag \Yu \Yd^\dag \Yd
\\&+ \Bigl [ 16 g_3^2 - \frac{2}{ 5} g_1^2 \Bigr] \Tr(\Yd \Yd^\dag)
  + \frac{6}{ 5} g_1^2 \Tr(\Ye \Ye^\dag)
  + \frac{4}{ 5} g_1^2 \Yu^\dag \Yu
  + \Bigl [6 g^2_2 + \frac{4}{ 5} g_1^2 \Bigr] \Yd^\dag \Yd 
\\&-\frac{16}{ 9} g_3^4 + 8 g_3^2 g_2^2 + \frac{8}{ 9} g_3^2 g_1^2
  + \frac{15}{ 2} g_2^4 + g_2^2 g_1^2 + \frac{287}{ 90} g_1^4
\biggr\rbrace\;,
\end{align*}
\begin{align*}
\beto_{\Yu} = \Yu \biggl\lbrace
&\Tr (3\Yu \Yu^\dag +\Yn \Yn^\dag) + 3 \Yu^\dag \Yu+ \Yd^\dag \Yd
- \frac{16}{ 3} g_3^2 - 3 g_2^2 - \frac{13}{ 15} g_1^2
\biggr \rbrace\;,
\eqnum\displaybreak[2]\\
\bett_{\Yu} =
\Yu \biggl\lbrace
&-\Tr (9\Yu \Yu^\dag \Yu \Yu^\dag +  3\Yu \Yd^\dag \Yd \Yu^\dag
  + 3\Yn \Yn^\dag \Yn \Yn^\dag + \Yn \Ye^\dag \Ye \Yn^\dag)
\eqnum\\&- \Yd^\dag \Yd \Tr (3 \Yd \Yd^\dag+ \Ye \Ye^\dag)
  - 3 \Yu^\dag \Yu \Tr (3\Yu \Yu^\dag +\Yn \Yn^\dag)
\\& - 4 \Yu^\dag \Yu \Yu^\dag \Yu - 2 \Yd^\dag \Yd \Yd^\dag \Yd
  - 2 \Yd^\dag \Yd \Yu^\dag \Yu
\\&+ \Bigl [ 16 g_3^2 + \frac{4}{ 5} g_1^2 \Bigr] \Tr(\Yu \Yu^\dag)
  + \Bigl [ 6 g_2^2 + \frac{2}{ 5} g_1^2 \Bigr ] \Yu^\dag \Yu
  + \frac{2}{ 5} g_1^2 \Yd^\dag \Yd
\\&-\frac{16}{ 9} g_3^4 + 8 g_3^2 g_2^2 + \frac{136}{ 45} g_3^2 g_1^2
  + \frac{15}{ 2} g_2^4 + g_2^2 g_1^2 + \frac{2743}{ 450} g_1^4
\biggr\rbrace\;,
\end{align*}
\begin{align*}
\beto_{\Ye} =\Ye \biggl\lbrace
&\Tr (3 \Yd \Yd^\dag + \Ye \Ye^\dag)  + 3 \Ye^\dag \Ye +\Yn^\dag \Yn
  - 3 g_2^2 - \frac{9}{ 5} g_1^2
\biggr \rbrace\;,
\eqnum\displaybreak[2]\\
\bett_{\Ye} =
\Ye \biggl\lbrace
&- \Tr (9\Yd \Yd^\dag \Yd \Yd^\dag + 3\Yu \Yd^\dag \Yd \Yu^\dag
  + 3\Ye \Ye^\dag \Ye \Ye^\dag + \Yn \Ye^\dag \Ye \Yn^\dag)
\eqnum\\&- \Yn^\dag \Yn \Tr(3 \Yu \Yu^\dag + \Yn \Yn^\dag)
  - 3\Ye^\dag \Ye \Tr(3 \Yd \Yd^\dag + \Ye \Ye^\dag)
\\&- 4 \Ye^\dag \Ye \Ye^\dag \Ye - 2 \Yn^\dag \Yn \Yn^\dag \Yn
  - 2 \Yn^\dag \Yn \Ye^\dag \Ye
\\&+ \Bigl [ 16 g_3^2 - \frac{2}{  5} g_1^2 \Bigr] \Tr(\Yd \Yd^\dag)
  + \frac{6}{  5} g_1^2 \Tr(\Ye \Ye^\dag) + 6 g_2^2 \Ye^\dag \Ye
\\&+ \frac{15}{ 2} g_2^4 + \frac{9}{ 5}g_2^2 g_1^2 + \frac{27}{ 2} g_1^4
\biggr\rbrace\;,
\end{align*}
\begin{align*}
\beto_{\Yn}  = \Yn \biggl\lbrace
&\Tr (3 \Yu \Yu^\dag + \Yn \Yn^\dag)
  + 3 \Yn^\dag \Yn +\Ye^\dag \Ye - 3 g_2^2 - \frac{3}{5} g_1^2
\biggr\rbrace\;,
\eqnum\label{beta1loopYn}\displaybreak[2]\\
\bett_{\Yn}  = 
\Yn \biggl\lbrace
&- \Tr (9\Yu \Yu^\dag \Yu \Yu^\dag  + 3\Yu \Yd^\dag \Yd \Yu^\dag
  + 3\Yn \Yn^\dag \Yn \Yn^\dag  + \Yn \Ye^\dag \Ye \Yn^\dag)
\eqnum \label{beta2loopYn}
\\&- \Ye^\dag \Ye \Tr(3 \Yd \Yd^\dag + \Ye \Ye^\dag)
  - 3\Yn^\dag \Yn \Tr(3 \Yu \Yu^\dag + \Yn \Yn^\dag)
\\&- 4 \Yn^\dag \Yn \Yn^\dag \Yn - 2 \Ye^\dag \Ye \Ye^\dag \Ye
  - 2 \Ye^\dag \Ye \Yn^\dag \Yn
\\&+ \Bigl[ 16 g_3^2 +\frac{4}{5} g_1^2 \Bigr] \Tr(\Yu \Yu^\dag)
  + \Bigl[6 g_2^2 + \frac{6}{5} g_1^2\Bigr] \Yn^\dag \Yn
  + \frac{6}{5} g_1^2 \Ye^\dag \Ye
\\&+ \frac{15}{ 2} g_2^4 + \frac{9}{ 5}g_2^2 g_1^2 + \frac{207}{50} g_1^4
\biggr\rbrace\;,
\end{align*}
\begin{align*}
\beto_\mu  = \mu \biggl \lbrace
&\Tr ( 3\Yu \Yu^\dag +3  \Yd \Yd^\dag + \Ye \Ye^\dag +\Yn \Yn^\dag)
  - 3 g_2^2 - \frac{3}{5} g_1^2 \biggr \rbrace\;,
\eqnum\displaybreak[2]\\
\bett_\mu  = 
\mu \biggl\lbrace
&- \Tr (\bad
    &9\Yu \Yu^\dag \Yu \Yu^\dag +9\Yd \Yd^\dag \Yd \Yd^\dag
      +6\Yu \Yd^\dag \Yd \Yu^\dag
   \\&
    +3\Ye \Ye^\dag \Ye \Ye^\dag +3\Yn \Yn^\dag \Yn \Yn^\dag
      +2 \Yn \Ye^\dag \Ye \Yn^\dag)\ead
\eqnum\\&+ \Bigl [16 g_3^2 + \frac{4}{5} g_1^2 \Bigr ]\Tr(\Yu \Yu^\dag )
  + \Bigl [16 g_3^2 - \frac{2}{5} g_1^2 \Bigr] \Tr(\Yd \Yd^\dag )
  + \frac{6}{5} g_1^2 \Tr(\Ye \Ye^\dag )
\\&+ \frac{15}{2} g_2^4 + \frac{9}{5} g_1^2 g_2^2 + \frac{207}{ 50} g_1^4
\biggr\rbrace\;,
\end{align*}
\begin{align*}
\beto_\sM = \sM \biggl [& 2 \Yn^\con \Yn^\tra \biggr ] + \biggl [2\Yn \Yn^\dag \biggr ] \sM
\;,
\eqnum\displaybreak[2]\\
\bett_\sM  = \sM \eqnum  \biggl [ 
&-2 \Yn^\con \Ye^\tra \Ye^\con \Yn^\tra -2 \Yn^\con \Yn^\tra \Yn^\con \Yn^\tra
  -2 \Yn^\con \Yn^\tra \Tr (3 \Yu \Yu^\dag + \Yn \Yn^\dag)
\\&+\frac{6}{5} g_1^2 \Yn^\con \Yn^\tra + 6 g_2^2 \Yn^\con \Yn^\tra
    \biggr ] +
\\ \biggl [
&-2 \Yn \Ye^\dag \Ye \Yn^\dag -2 \Yn \Yn^\dag \Yn \Yn^\dag
   -2 \Yn \Yn^\dag \Tr (3 \Yu \Yu^\dag + \Yn \Yn^\dag)
\\&+\frac{6}{5} g_1^2 \Yn \Yn^\dag + 6 g_2^2 \Yn \Yn^\dag
\biggr ] \sM\;.
\end{align*}

The soft SUSY breaking Lagrangian of the MSSM extended with
right-handed neutrinos reads:
\begin{eqnarray}  
-{\cal L}_{\rm soft}&=&\  
\mHus H^*_u H_u+\mHds H^*_d H_d +
(\mQs)_{ij} \widetilde Q^*_i  \widetilde Q_j+
(\mds)_{ij} \widetilde d^*_{Ri}  \widetilde d_{Rj}+
(\mus)_{ij} \widetilde u^*_{Ri}  \widetilde u_{Rj}+\nonumber \\
&&
(\mLs)_{ij} \widetilde L^*_i  \widetilde L_j\ +
(\mes)_{ij} \widetilde e^*_{Ri}  \widetilde e_{Rj}\ + 
(\mns)_{ij} \widetilde \nu^*_{Ri}  \widetilde \nu_{Rj}\ +  
\left(M_1 \widetilde B \widetilde B 
+M_2 \widetilde W\widetilde W +M_3 \widetilde g\widetilde g \right.
\nonumber \\
&&\left.
 \Ad_{ij} \widetilde d^*_{Ri} H_d \widetilde Q_j +
\Au_{ij} \widetilde u^*_{Ri} H_u \widetilde Q_j +
\Ae_{ij} \widetilde e^*_{Ri} H_d \widetilde L_j +
\An_{ij} \widetilde \nu^*_{Ri} H_u \widetilde L_j + \right. \nonumber \\
&&\left.
B H_u H_d - \frac{1}{2}\widetilde\nu_{Ri} \BM_{ij}\widetilde \nu_{Rj}+
{\rm h.c.}\right)\;.
\end{eqnarray}

The $\beta$-functions of the soft gaugino masses are given by
\begin{align}
\beto_{M_a} &= 2 g_a^2 B^{(1)}_a M_a\;,\\ 
\bett_{M_a} &= 2 g_a^2 \biggl [\sum_{b=1}^3 B^{(2)}_{ab} g_b^2 (M_a + M_b)
+ \sum_{x=u,d,e,\nu} C_a^x \left ( \Tr [\Yx^\dag \Ax] -
M_a \Tr[\Yx^\dag \Yx] \right ) \biggr ]\;,
\end{align}
where $B^{(1)}_a$, $B^{(2)}_{ab}$ and $C^{u,d,e,\nu}_a$ were
defined in Eq.~(\ref{defs-gauge}). On the other hand, the
$\beta$-functions of the soft scalar masses read:
\begin{align*}
\beto_{\mHus} = \:
&\Tr [ \eqnum\bad
  &6(\mHus + \mQs)\Yu^\dag \Yu + 6\Yu^\dag \mus \Yu + 6\Au^\dag \Au
     +2(\mHus + \mLs)\Yn^\dag \Yn 
  \\&+ 2\Yn^\dag \mns \Yn + 2\An^\dag \An ]
    - 6 g_2^2 |M_2|^2 - \frac{6}{5} g_1^2 |M_1|^2
    + \frac{3}{5} g_1^2 \trym \;,\ead
\displaybreak[2]\\
\bett_{\mHus} = \:
&-2 \Tr \biggl [ \eqnum \bad
  &+18 (\mHus + \mQs) \Yu^\dag \Yu \Yu^\dag \Yu
    + 18 \Yu^\dag \mus \Yu \Yu^\dag \Yu
  \\&+ 3(\mHus+\mHds+\mQs) \Yu^\dag \Yu \Yd^\dag \Yd
    + 3\Yu^\dag \mus \Yu \Yd^\dag \Yd
  \\&+ 3\Yu^\dag  \Yu \mQs\Yd^\dag \Yd +  3\Yu^\dag  \Yu \Yd^\dag \mds \Yd
  \\&+6 (\mHus + \mLs) \Yn^\dag \Yn \Yn^\dag \Yn 
    + 6 \Yn^\dag \mns \Yn \Yn^\dag \Yn 
  \\&+ (\mHus+\mHds+\mLs) \Yn^\dag \Yn \Ye^\dag \Ye
    + \Yn^\dag \mns \Yn \Ye^\dag \Ye 
  \\&+ \Yn^\dag  \Yn \mLs \Ye^\dag \Ye
     + \Yn^\dag  \Yn \Ye^\dag \mes \Ye 
     +18 \Au^\dag \Au \Yu^\dag \Yu
  \\& +18 \Au^\dag \Yu \Yu^\dag \Au
    + 3\Ad^\dag \Ad \Yu^\dag \Yu + 3\Yd^\dag \Yd \Au^\dag \Au  
  \\&+\!3\Ad^\dag \Yd \Yu^\dag \Au +\!3\Yd^\dag \Ad \Au^\dag \Yu 
    +\!6 \An^\dag \An \Yn^\dag \Yn +\!6 \An^\dag \Yn \Yn^\dag \An 
  \\&+ \Ae^\dag \Ae \Yn^\dag \Yn + \Ye^\dag \Ye \An^\dag \An
    + \Ae^\dag \Ye \Yn^\dag \An 
    + \Ye^\dag \Ae \An^\dag \Yn \biggr] 
\ead \\&+\Bigl [ 32 g_3^2 + \frac{8}{5}g_1^2 \Bigr ]
  \Tr [(\mHus + \mQs)  \Yu^\dag \Yu + \Yu^\dag \mus \Yu + \Au^\dag \Au ] 
\\&+ 32 g_3^2 \Bigl \lbrace 
  2 |M_3|^2 \Tr [\Yu^\dag \Yu] - M_3^* \Tr [ \Yu^\dag\Au  ] - M_3 \Tr [\Au^\dag \Yu  ]
  \Bigr \rbrace 
\\& + \frac{8}{5} g_1^2 \Bigl \lbrace
  2 |M_1|^2 \Tr [\Yu^\dag \Yu] - M_1^* \Tr [ \Yu^\dag \Au ] - M_1 \Tr [  \Au^\dag\Yu ]
  \Bigr \rbrace
  + \frac{6}{5} g_1^2  \sss 
\\&+ 33 g_2^4 |M_2|^2 + \frac{18}{5} g_2^2 g_1^2 (|M_2|^2 + |M_1|^2 + \Re[M_1 M_2^*])
  +\frac{621}{25} g_1^4 |M_1|^2 
\\&+ 3 g_2^2 \sigma_2 + \frac{3}{5} g_1^2 \sigma_1\;,
\end{align*}
\begin{align*}
\beto_{\mHds} =\:
&\Tr \Bigl [ \eqnum \bad
  &6(\mHds + \mQs) \Yd^\dag \Yd + 6\Yd^\dag \mds  \Yd
    +2(\mHds + \mLs)\Ye^\dag \Ye + 2 \Ye^\dag \mes \Ye
  \\&+ 6 \Ad^\dag \Ad + 2 \Ae^\dag \Ae \Bigr]
  - 6 g_2^2 |M_2|^2 - \frac{6}{5} g_1^2 |M_1|^2
  - \frac{3}{5} g_1^2 \trym\;,
\ead \displaybreak[2]\\
\bett_{\mHds} = \:
&-2 \Tr \biggl [ \eqnum \bad
  &+18(\mHds+\mQs) \Yd^\dag \Yd \Yd^\dag \Yd
    + 18 \Yd^\dag \mds \Yd \Yd^\dag \Yd
  \\&+ 3(\mHus+\mHds + \mQs) \Yu^\dag \Yu \Yd^\dag \Yd
    + 3\Yu^\dag \mus \Yu \Yd^\dag \Yd
  \\&+ 3\Yu^\dag  \Yu \mQs\Yd^\dag \Yd
    + 3\Yu^\dag  \Yu \Yd^\dag \mds \Yd
  \\&+ 6 (\mHds + \mLs)\Ye^\dag \Ye \Ye^\dag \Ye
    + 6 \Ye^\dag \mes \Ye \Ye^\dag \Ye
  \\&+ (\mHus+\mHds + \mLs) \Yn^\dag \Yn \Ye^\dag \Ye
     + \Yn^\dag \mns \Yn \Ye^\dag \Ye
  \\&+ \Yn^\dag  \Yn \mLs \Ye^\dag \Ye
     + \Yn^\dag  \Yn \Ye^\dag \mes \Ye 
     + 18 \Ad^\dag \Ad \Yd^\dag \Yd
  \\& + 18 \Ad^\dag \Yd \Yd^\dag \Ad
    + 3\Au^\dag \Au \Yd^\dag \Yd + 3\Yu^\dag \Yu \Ad^\dag \Ad 
  \\&+ 3\Au^\dag \Yu \Yd^\dag \Ad +3\Yu^\dag \Au \Ad^\dag \Yd
    + 6 \Ae^\dag \Ae \Ye^\dag \Ye + 6 \Ae^\dag \Ye \Ye^\dag \Ae 
  \\&+ \An^\dag \An \Ye^\dag \Ye 
    + \Yn^\dag \Yn \Ae^\dag \Ae
    +  \An^\dag \Yn \Ye^\dag \Ae
    +  \Yn^\dag \An \Ae^\dag \Ye  \biggr ] 
\ead \\&+\Bigl [ 32 g_3^2 - \frac{4}{5}g_1^2 \Bigr ]
  \Tr [ (\mHds + \mQs)\Yd^\dag \Yd + \Yd^\dag \mds \Yd + \Ad^\dag \Ad ] 
\\& + 32 g_3^2 \Bigl \lbrace 2 |M_3|^2 \Tr [\Yd^\dag \Yd]
  - M_3^* \Tr [ \Yd^\dag\Ad  ] - M_3 \Tr [\Ad^\dag \Yd  ]
  \Bigr \rbrace 
\\& - \frac{4}{5} g_1^2 \Bigl \lbrace 2 |M_1|^2 \Tr [\Yd^\dag \Yd]
  - M_1^* \Tr [ \Yd^\dag \Ad ] - M_1 \Tr [  \Ad^\dag\Yd ]
  \Bigr \rbrace 
\\&+ \frac{12}{5} g_1^2 \Bigl \lbrace \bad
    &\Tr [ (\mHds + \mLs)\Ye^\dag\Ye +  \Ye^\dag \mes \Ye +\Ae^\dag \Ae ] 
    \\&+2 |M_1|^2 \Tr[\Ye^\dag \Ye ] - M_1 \Tr[\Ae^\dag \Ye ]
      - M^*_1 \Tr[\Ye^\dag \Ae ] \Bigr \rbrace
\ead \\&- \frac{6}{5} g_1^2  \sss + 33 g_2^4 |M_2|^2
  + \frac{18}{5} g_2^2 g_1^2 (|M_2|^2 + |M_1|^2 + \Re[M_1 M_2^*]) 
\\&+\frac{621}{25} g_1^4 |M_1|^2 
  + 3 g_2^2 \sigma_2 + \frac{3}{5} g_1^2 \sigma_1\;,
\end{align*}
\begin{align*}
\beto_{\mQs} = \:
& (\mQs + 2 \mHus ) \Yu^\dag \Yu + (\mQs + 2 \mHds ) \Yd^\dag \Yd
   + 2 \Yu^\dag \mus \Yu + 2 \Yd^\dag \mds \Yd
\eqnum\\& + [ \Yu^\dag \Yu + \Yd^\dag \Yd ] \mQs
  + 2 \Au^\dag \Au + 2 \Ad^\dag \Ad
\\&- \frac{32}{3} g_3^2 |M_3|^2 - 6 g_2^2 |M_2|^2 - \frac{2}{15} g_1^2 |M_1|^2
  + \frac{1}{5} g_1^2 \trym\;,
\displaybreak[2]\\
\bett_{\mQs} =
& -(2 \mQs + 8 \mHus ) \Yu^\dag \Yu\Yu^\dag \Yu
  -4  \Yu^\dag \mus \Yu\Yu^\dag \Yu
  -4  \Yu^\dag  \Yu \mQs  \Yu^\dag \Yu
\eqnum\\
&-4  \Yu^\dag  \Yu \Yu^\dag \mus \Yu 
  -2  \Yu^\dag  \Yu \Yu^\dag \Yu \mQs
  -(2 \mQs + 8 \mHds ) \Yd^\dag \Yd\Yd^\dag \Yd
\\&-4  \Yd^\dag \mds \Yd\Yd^\dag \Yd
  -4  \Yd^\dag  \Yd \mQs  \Yd^\dag \Yd
  -4  \Yd^\dag  \Yd \Yd^\dag \mds \Yd
  -2  \Yd^\dag  \Yd \Yd^\dag \Yd \mQs 
\\&- \Bigl [ (\mQs + 4 \mHus )\Yu^\dag \Yu
  + 2 \Yu^\dag \mus \Yu +  \Yu^\dag \Yu \mQs
  \Bigr ] \Tr(3 \Yu^\dag \Yu +\Yn^\dag \Yn) 
\\&- \Bigl [ (\mQs + 4 \mHds )\Yd^\dag \Yd
  + 2 \Yd^\dag \mds \Yd +  \Yd^\dag \Yd \mQs
  \Bigr ] \Tr(3 \Yd^\dag \Yd + \Ye^\dag \Ye ) 
\\& - \Yu^\dag \Yu \Tr ( 6\mQs\Yu^\dag \Yu  + 6\Yu^\dag \mus \Yu 
  +2\mLs \Yn^\dag \Yn  +  2 \Yn^\dag \mns \Yn) 
\\& - \Yd^\dag \Yd \Tr (6 \mQs\Yd^\dag \Yd  +
  6 \Yd^\dag \mds \Yd + 2\mLs \Ye^\dag \Ye +
  2\Ye^\dag \mes \Ye ) 
\\&- 4 \Bigl \lbrace \Yu^\dag \Yu \Au^\dag \Au
  + \Au^\dag \Au \Yu^\dag \Yu + \Yu^\dag \Au \Au^\dag \Yu
  + \Au^\dag \Yu \Yu^\dag \Au   \Bigr\rbrace 
\\& - 4 \Bigl\lbrace \Yd^\dag \Yd \Ad^\dag \Ad + \Ad^\dag \Ad \Yd^\dag \Yd
  + \Yd^\dag \Ad \Ad^\dag \Yd + \Ad^\dag \Yd \Yd^\dag \Ad    \Bigr\rbrace 
\\&- \Au^\dag \Au  \Tr[ 6 \Yu^\dag \Yu +2 \Yn^\dag \Yn]
  - \Yu^\dag \Yu  \Tr[ 6 \Au^\dag \Au +2 \An^\dag \An ] 
\\&- \Au^\dag \Yu  \Tr[ 6 \Yu^\dag \Au +2 \Yn^\dag \An]
  - \Yu^\dag \Au  \Tr[ 6 \Au^\dag \Yu +2 \An^\dag \Yn] 
\\&- \Ad^\dag \Ad  \Tr[ 6 \Yd^\dag \Yd + 2 \Ye^\dag \Ye ]
  - \Yd^\dag \Yd  \Tr[ 6 \Ad^\dag \Ad + 2 \Ae^\dag \Ae ] 
\\&- \Ad^\dag \Yd  \Tr[ 6 \Yd^\dag \Ad + 2 \Ye^\dag \Ae ]
  - \Yd^\dag \Ad  \Tr[ 6 \Ad^\dag \Yd + 2 \Ae^\dag \Ye ] 
\\&+ \frac{2}{5} g_1^2 \biggl \lbrace \bad
    &(2\mQs + 4 \mHus )\Yu^\dag \Yu + 4 \Yu^\dag \mus \Yu
      + 2 \Yu^\dag \Yu \mQs + 4 \Au^\dag \Au - 4 M_1\Au^\dag \Yu 
    \\&- 4 M_1^* \Yu^\dag \Au  + 8 |M_1|^2 \Yu^\dag \Yu
      +(\mQs + 2 \mHds )\Yd^\dag \Yd + 2 \Yd^\dag \mds \Yd 
    \\&+  \Yd^\dag \Yd \mQs + 2 \Ad^\dag \Ad
      - 2 M_1\Ad^\dag \Yd - 2 M_1^* \Yd^\dag \Ad  + 4 |M_1|^2 \Yd^\dag \Yd
      \biggr \rbrace 
\ead \\&+ \frac{2}{5} g_1^2 \sss - \frac{128}{3} g_3^4 |M_3|^2
  + 32 g_3^2 g_2^2 (|M_3|^2 + |M_2|^2 + \Re[M_2 M_3^*]) 
\\&+ \frac{32}{45} g_3^2 g_1^2 (|M_3|^2 + |M_1|^2 + \Re[M_1 M_3^*])
  + 33 g_2^4 |M_2|^2 
\\&+ \frac{2}{5} g_2^2 g_1^2 (|M_2|^2 + |M_1|^2 + \Re[M_1 M_2^*])
  +\frac{199}{75} g_1^4 |M_1|^2 
\\& + \frac{16}{3} g_3^2 \sigma_3 + 3 g_2^2 \sigma_2 
  + \frac{1}{15} g_1^2\sigma_1\;,
\end{align*}
\begin{align*}
\beto_{\mds}  = \:
& (2\mds + 4 \mHds ) \Yd \Yd^\dag
  + 4\Yd \mQs\Yd^\dag + 2\Yd \Yd^\dag  \mds + 4 \Ad \Ad^\dag
\eqnum\\&- \frac{32}{3} g_3^2 |M_3|^2 - \frac{8}{15} g_1^2 |M_1|^2
  + \frac{2}{5} g_1^2 \trym\;,
\displaybreak[2]\\
\bett_{\mds} = \:
& -(2 \mds + 8 \mHds ) \Yd \Yd^\dag\Yd \Yd^\dag
  -4 \Yd \mQs \Yd^\dag\Yd \Yd^\dag
  -4 \Yd  \Yd^\dag \mds \Yd \Yd^\dag
\eqnum\\ &-4 \Yd  \Yd^\dag  \Yd \mQs \Yd^\dag
  -2 \Yd \Yd^\dag  \Yd \Yd^\dag \mds
  -(2 \mds + 4 \mHus + 4 \mHds) \Yd \Yu^\dag\Yu \Yd^\dag
\\&-4 \Yd \mQs \Yu^\dag\Yu \Yd^\dag
  -4 \Yd  \Yu^\dag \mus \Yu \Yd^\dag
  -4 \Yd  \Yu^\dag  \Yu \mQs \Yd^\dag
  -2 \Yd  \Yu^\dag  \Yu \Yd^\dag \mds
\\& - \Bigl [ (\mds+ 4 \mHds)
  \Yd \Yd^\dag + 2 \Yd \mQs \Yd^\dag
  + \Yd \Yd^\dag \mds \Bigr ]
  \Tr(6\Yd^\dag \Yd+2\Ye^\dag \Ye )
\\&- 4 \Yd \Yd^\dag  \Tr(3\mQs\Yd^\dag \Yd  + 3 \Yd^\dag \mds \Yd
  + \mLs \Ye^\dag \Ye  + \Ye^\dag \mes \Ye )
\\&- 4 \Bigl\lbrace \Ad \Ad^\dag \Yd \Yd^\dag
     +\Yd \Yd^\dag \Ad \Ad^\dag
     +\Ad \Yd^\dag \Yd \Ad^\dag
     +\Yd \Ad^\dag \Ad \Yd^\dag  \Bigr\rbrace
\\&- 4 \Bigl\lbrace \Ad \Au^\dag \Yu \Yd^\dag
     +\Yd \Yu^\dag \Au \Ad^\dag
     +\Ad \Yu^\dag \Yu \Ad^\dag
     +\Yd \Au^\dag \Au \Yd^\dag  \Bigr\rbrace
\\& - 4\Ad \Ad^\dag \Tr ( 3\Yd^\dag \Yd + \Ye^\dag \Ye )
  -4   \Yd \Yd^\dag \Tr ( 3\Ad^\dag \Ad +\Ae^\dag \Ae )
\\&-4  \Ad \Yd^\dag \Tr ( 3 \Ad^\dag \Yd + \Ae^\dag \Ye )
  -4 \Yd \Ad^\dag \Tr ( 3 \Yd^\dag \Ad + \Ye^\dag \Ae )
\\&+ \Bigl [ 6 g_2^2 + \frac{2}{5} g_1^2 \Bigr ] \Bigl \lbrace (\mds+ 2 \mHds)
  \Yd\Yd^\dag + 2 \Yd \mQs\Yd^\dag
  + \Yd\Yd^\dag \mds + 2 \Ad \Ad^\dag \Bigr\rbrace
\\& + 12 g_2^2 \Bigl \lbrace 2 |M_2|^2 \Yd \Yd^\dag -
  M_2^* \Ad \Yd^\dag - M_2 \Yd \Ad^\dag \Bigr \rbrace
\\&+ \frac{4}{5} g_1^2 \Bigl \lbrace 2 |M_1|^2 \Yd \Yd^\dag -
  M_1^* \Ad \Yd^\dag - M_1 \Yd \Ad^\dag \Bigr \rbrace
  + \frac{4}{5} g_1^2 \sss
\\&- \frac{128}{3} g_3^4 |M_3|^2
  + \frac{128}{45} g_3^2 g_1^2 (|M_3|^2 + |M_1|^2 + \Re[M_1 M_3^*])
  +\frac{808}{75} g_1^4 |M_1|^2
\\ &+ \frac{16}{3} g_3^2 \sigma_3 + \frac{4}{15} g_1^2 \sigma_1\;,
\end{align*}
\begin{align*}
\beto_{\mus} = \:
&(2\mus + 4 \mHus ) \Yu \Yu^\dag
  + 4\Yu \mQs\Yu^\dag
  + 2\Yu \Yu^\dag  \mus
  + 4 \Au \Au^\dag
\eqnum\\ & - \frac{32}{3} g_3^2 |M_3|^2 - \frac{32}{15} g_1^2 |M_1|^2
  - \frac{4}{5} g_1^2 \trym\;,
\displaybreak[2]\\
\bett_{\mus} = \:
& -(2 \mus + 8 \mHus ) \Yu \Yu^\dag\Yu \Yu^\dag
  -4 \Yu \mQs \Yu^\dag\Yu \Yu^\dag
  -4 \Yu  \Yu^\dag \mus \Yu \Yu^\dag
\eqnum\\ & -4 \Yu  \Yu^\dag  \Yu \mQs \Yu^\dag
  -2 \Yu  \Yu^\dag  \Yu \Yu^\dag \mus
  -(2 \mus + 4 \mHus + 4 \mHds) \Yu \Yd^\dag\Yd \Yu^\dag
\\&-4 \Yu \mQs \Yd^\dag\Yd \Yu^\dag
  -4 \Yu  \Yd^\dag \mds \Yd \Yu^\dag
  -4 \Yu  \Yd^\dag  \Yd \mQs \Yu^\dag
  -2 \Yu  \Yd^\dag  \Yd \Yu^\dag \mus
\\& - \Bigl [ (\mus+ 4 \mHus) \Yu \Yu^\dag + 2 \Yu \mQs \Yu^\dag
  + \Yu \Yu^\dag \mus \Bigr ]\Tr[6\Yu^\dag \Yu  + 2\Yn^\dag \Yn]
\\&- 4 \Yu \Yu^\dag \Tr[3\mQs\Yu^\dag \Yu  + 3\Yu^\dag \mus \Yu 
  +\mLs\Yn^\dag \Yn  + \Yn^\dag \mns \Yn]
\\&- 4\left \lbrace \Au \Au^\dag \Yu \Yu^\dag
     +\Yu \Yu^\dag \Au \Au^\dag
     +\Au \Yu^\dag \Yu \Au^\dag
     +\Yu \Au^\dag \Au \Yu^\dag  \right \rbrace
\\&- 4\left\lbrace \Au \Ad^\dag \Yd \Yu^\dag
     +\Yu \Yd^\dag \Ad \Au^\dag
     +\Au \Yd^\dag \Yd \Au^\dag
     +\Yu \Ad^\dag \Ad \Yu^\dag  \right \rbrace
\\& -4 \Au \Au^\dag \Tr ( 3\Yu^\dag \Yu + \Yn^\dag \Yn)
  -4 \Yu \Yu^\dag \Tr ( 3\Au^\dag \Au+ \An^\dag \An)
\\ & -4 \Au \Yu^\dag \Tr ( 3\Au^\dag \Yu + \An^\dag \Yn)
  -4 \Yu \Au^\dag \Tr ( 3\Yu^\dag \Au + \Yn^\dag \An)
\\&+ \Bigl [ 6 g_2^2 - \frac{2}{5} g_1^2 \Bigr ] 
  \Bigl \lbrace (\mus+ 2 \mHus) \Yu\Yu^\dag + 2 \Yu \mQs\Yu^\dag
  + \Yu\Yu^\dag \mus + 2 \Au \Au^\dag \Bigr \rbrace
\\& + 12 g_2^2 \Bigl \lbrace 2 |M_2|^2 \Yu \Yu^\dag -
  M_2^* \Au \Yu^\dag - M_2 \Yu \Au^\dag \Bigr \rbrace
\\&- \frac{4}{5} g_1^2 \Bigl \lbrace 2 |M_1|^2 \Yu \Yu^\dag -
  M_1^* \Au \Yu^\dag - M_1 \Yu \Au^\dag \Bigr \rbrace
- \frac{8}{5} g_1^2 \sss
\\& - \frac{128}{3} g_3^4 |M_3|^2
  + \frac{512}{45} g_3^2 g_1^2 (|M_3|^2 + |M_1|^2 + \Re[M_1 M_3^*])
  +\frac{3424}{75} g_1^4 |M_1|^2
\\ &+ \frac{16}{3} g_3^2 \sigma_3 + \frac{16}{15} g_1^2 \sigma_1\;,
\end{align*}
\begin{align*}
\beto_{\mLs} = \:
&+ (\mLs + 2 \mHus ) \Yn^\dag \Yn + (\mLs + 2 \mHds ) \Ye^\dag \Ye
  + 2 \Yn^\dag \mns \Yn + 2 \Ye^\dag \mes \Ye
\eqnum \label{beta1loop-mLs}\\
&+[\Yn^\dag \Yn +\Ye^\dag \Ye]\mLs + 2 \An^\dag \An + 2 \Ae^\dag \Ae 
\\&- 6 g_2^2 |M_2|^2- \frac{6}{5} g_1^2 |M_1|^2
  - \frac{3}{5} g_1^2 \trym\;,
\displaybreak[2]\\
\bett_{\mLs} = \:
& -(2 \mLs + 8 \mHus ) \Yn^\dag \Yn\Yn^\dag \Yn
  -4  \Yn^\dag \mns \Yn\Yn^\dag \Yn
  -4  \Yn^\dag  \Yn \mLs \Yn^\dag \Yn
\eqnum\label{beta2loop-mLs} \\ &
  -4  \Yn^\dag \Yn\Yn^\dag \mns \Yn
  -2  \Yn^\dag \Yn\Yn^\dag  \Yn \mLs
  -(2 \mLs + 8 \mHds ) \Ye^\dag \Ye\Ye^\dag \Ye
\\ & 
  -4  \Ye^\dag \mes \Ye\Ye^\dag \Ye
  -4  \Ye^\dag  \Ye \mLs \Ye^\dag \Ye
  -4  \Ye^\dag \Ye\Ye^\dag \mes \Ye
  -2  \Ye^\dag \Ye\Ye^\dag  \Ye \mLs
\\ &
  - \Bigl [ (\mLs + 4 \mHus )\Yn^\dag \Yn + 2 \Yn^\dag \mns \Yn +  \Yn^\dag \Yn \mLs     \Bigr ]
  \Tr (3 \Yu^\dag \Yu + \Yn^\dag \Yn )
\\ &
  - \Bigl [ (\mLs + 4 \mHds )\Ye^\dag \Ye + 2 \Ye^\dag \mes \Ye +\Ye^\dag \Ye \mLs \Bigr ]
  \Tr (3 \Yd^\dag \Yd + \Ye^\dag \Ye )
\\ &
  - \Yn^\dag \Yn \Tr [6 \mQs \Yu^\dag \Yu+6\Yu^\dag \mus \Yu + 2\mLs\Yn^\dag \Yn +     2\Yn^\dag \mns \Yn ]
\\ &
  - \Ye^\dag \Ye \Tr [6 \mQs \Yd^\dag \Yd  + 6\Yd^\dag \mds \Yd + 2\mLs\Ye^\dag \Ye + 2\Ye^\dag \mes \Ye ]
\\ &
  - 4 \Bigl\lbrace \Yn^\dag \Yn \An^\dag \An + \An^\dag \An \Yn^\dag \Yn
  + \Yn^\dag \An \An^\dag \Yn + \An^\dag \Yn \Yn^\dag \An \Bigr\rbrace
\\ &
  - 4 \Bigl\lbrace \Ye^\dag \Ye \Ae^\dag \Ae + \Ae^\dag \Ae \Ye^\dag \Ye
  + \Ye^\dag \Ae \Ae^\dag \Ye + \Ae^\dag \Ye \Ye^\dag \Ae \Bigr\rbrace
\\ &
  - \An^\dag \An  \Tr[ 6 \Yu^\dag \Yu + 2 \Yn^\dag \Yn ]
  - \Yn^\dag \Yn  \Tr[ 6 \Au^\dag \Au + 2 \An^\dag \An ]
\\ &
  - \An^\dag \Yn  \Tr[ 6 \Yu^\dag \Au +2 \Yn^\dag \An ]
  - \Yn^\dag \An  \Tr[ 6 \Au^\dag \Yu + 2 \An^\dag \Yn ]
\\ &
  - \Ae^\dag \Ae  \Tr[ 6 \Yd^\dag \Yd + 2 \Ye^\dag \Ye ]
  - \Ye^\dag \Ye  \Tr[ 6 \Ad^\dag \Ad + 2 \Ae^\dag \Ae ]
\\ &
  - \Ae^\dag \Ye  \Tr[ 6 \Yd^\dag \Ad + 2 \Ye^\dag \Ae ]
  - \Ye^\dag \Ae  \Tr[ 6 \Ad^\dag \Yd + 2 \Ae^\dag \Ye ]
\\ &
+ \frac{6}{5} g_1^2 \biggl \lbrace \bad
  & (\mLs + 2 \mHds )\Ye^\dag \Ye + 2 \Ye^\dag \mes \Ye
    + \Ye^\dag \Ye \mLs + 2 \Ae^\dag \Ae
  \\ & - 2 M_1  \Ae^\dag \Ye - 2 M_1^* \Ye^\dag \Ae 
    + 4 |M_1|^2 \Ye^\dag \Ye \biggr \rbrace \ead
\\ & - \frac{6}{5} g_1^2 \sss + 33 g_2^4 |M_2|^2
  + \frac{18}{5} g_2^2 g_1^2 (|M_2|^2 + |M_1|^2 + \Re[M_1 M_2^*])
\\ &+\frac{621}{25} g_1^4 |M_1|^2
  + 3 g_2^2 \sigma_2 + \frac{3}{5} g_1^2 \sigma_1\;,
\end{align*}
\begin{align*}
\beto_{\mes} = \:
& (2\mes + 4 \mHds ) \Ye \Ye^\dag + 4\Ye \mLs\Ye^\dag
  + 2\Ye \Ye^\dag  \mes + 4 \Ae \Ae^\dag
\eqnum\\&- \frac{24}{5} g_1^2 |M_1|^2
  + \frac{6}{5} g_1^2 \trym\;,
\displaybreak[2]\\
\bett_{\mes}  = \:
& -(2 \mes + 8 \mHds ) \Ye \Ye^\dag\Ye \Ye^\dag
  -4 \Ye \mLs \Ye^\dag\Ye \Ye^\dag -4 \Ye  \Ye^\dag \mes \Ye \Ye^\dag
\eqnum\\ &-4 \Ye  \Ye^\dag  \Ye \mLs \Ye^\dag -2 \Ye \Ye^\dag  \Ye \Ye^\dag \mes
  -(2 \mes + 4 \mHus + 4 \mHds) \Ye \Yn^\dag\Yn \Ye^\dag
\\& -4 \Ye \mLs \Yn^\dag\Yn \Ye^\dag -4 \Ye  \Yn^\dag \mns \Yn \Ye^\dag
  -4 \Ye  \Yn^\dag  \Yn \mLs \Ye^\dag -2 \Ye  \Yn^\dag  \Yn \Ye^\dag \mes
\\& - \Bigl [ (\mes+ 4 \mHds) \Ye \Ye^\dag + 2 \Ye \mLs \Ye^\dag
  + \Ye \Ye^\dag \mes \Bigr ] \Tr(6\Yd^\dag \Yd+2\Ye^\dag \Ye )
\\&- 4 \Ye \Ye^\dag \Tr(3\mQs\Yd^\dag \Yd  + 3 \Yd^\dag \mds \Yd
  + \mLs \Ye^\dag \Ye  + \Ye^\dag \mes \Ye )
\\&- 4 \Bigl\lbrace \Ae \Ae^\dag \Ye \Ye^\dag
     +\Ye \Ye^\dag \Ae \Ae^\dag
     +\Ae \Ye^\dag \Ye \Ae^\dag
     +\Ye \Ae^\dag \Ae \Ye^\dag  \Bigr\rbrace
\\&- 4 \Bigl\lbrace \Ae \An^\dag \Yn \Ye^\dag
     +\Ye \Yn^\dag \An \Ae^\dag
     +\Ae \Yn^\dag \Yn \Ae^\dag
     +\Ye \An^\dag \An \Ye^\dag  \Bigr\rbrace
\\& - 4\Ae \Ae^\dag \Tr ( 3\Yd^\dag \Yd + \Ye^\dag \Ye )
  -4   \Ye \Ye^\dag \Tr ( 3\Ad^\dag \Ad + \Ae^\dag \Ae )
\\&-4  \Ae \Ye^\dag \Tr ( 3 \Ad^\dag \Yd + \Ae^\dag \Ye )
  -4 \Ye \Ae^\dag \Tr ( 3 \Yd^\dag \Ad + \Ye^\dag \Ae )
\\&+ \Bigl [ 6 g_2^2 - \frac{6}{5} g_1^2 \Bigr ] \Bigl \lbrace
  (\mes+ 2 \mHds)\Ye\Ye^\dag + 2 \Ye \mLs\Ye^\dag
  + \Ye\Ye^\dag \mes + 2 \Ae \Ae^\dag \Bigr\rbrace
\\& + 12 g_2^2 \Bigl \lbrace 2 |M_2|^2 \Ye \Ye^\dag -
  M_2^* \Ae \Ye^\dag - M_2 \Ye \Ae^\dag \Bigr \rbrace
\\& -\frac{12}{5} g_1^2 \Bigl \lbrace 2 |M_1|^2 \Ye \Ye^\dag -
  M_1^* \Ae \Ye^\dag - M_1 \Ye \Ae^\dag \Bigr \rbrace
\\&+ \frac{12}{5} g_1^2  \sss
  + \frac{2808}{25} g_1^4 |M_1|^2
  + \frac{12}{5} g_1^2 \sigma_1\;,
\end{align*}
\begin{align*}
\beto_{\mns} = \:
& (2\mns + 4 \mHus ) \Yn \Yn^\dag + 4\Yn \mLs\Yn^\dag
  + 2\Yn \Yn^\dag  \mns + 4 \An \An^\dag\;,
\eqnum\displaybreak[2]\\
\bett_{\mns} = \:
& -(2 \mns + 8 \mHus ) \Yn \Yn^\dag\Yn \Yn^\dag
  -4 \Yn \mLs \Yn^\dag\Yn \Yn^\dag -4 \Yn  \Yn^\dag \mns \Yn \Yn^\dag
\eqnum\\ & -4 \Yn  \Yn^\dag  \Yn \mLs \Yn^\dag -2 \Yn  \Yn^\dag  \Yn \Yn^\dag \mns
  -(2 \mns + 4 \mHus + 4 \mHds) \Yn \Ye^\dag\Ye \Yn^\dag
\\&-4 \Yn \mLs \Ye^\dag\Ye \Yn^\dag -4 \Yn  \Ye^\dag \mes \Ye \Yn^\dag
  -4 \Yn  \Ye^\dag  \Ye \mLs \Yn^\dag -2 \Yn  \Ye^\dag  \Ye \Yn^\dag \mns
\\& - \Bigl [ (\mns+ 4 \mHus) \Yn \Yn^\dag + 2 \Yn \mLs \Yn^\dag
  + \Yn \Yn^\dag \mns \Bigr ] \Tr[6\Yu^\dag \Yu  + 2\Yn^\dag \Yn]
\\&- 4 \Yn \Yn^\dag \Tr[3 \mQs\Yu^\dag \Yu  + 3 \Yu^\dag \mus \Yu 
  +\mLs\Yn^\dag \Yn  + \Yn^\dag \mns \Yn]
\\&- 4\left \lbrace \An \An^\dag \Yn \Yn^\dag
     +\Yn \Yn^\dag \An \An^\dag
     +\An \Yn^\dag \Yn \An^\dag
     +\Yn \An^\dag \An \Yn^\dag  \right \rbrace
\\&- 4\left\lbrace \An \Ae^\dag \Ye \Yn^\dag
     +\Yn \Ye^\dag \Ae \An^\dag
     +\An \Ye^\dag \Ye \An^\dag
     +\Yn \Ae^\dag \Ae \Yn^\dag  \right \rbrace
\\& -4 \An \An^\dag \Tr ( 3\Yu^\dag \Yu + \Yn^\dag \Yn)
  -4\Yn \Yn^\dag \Tr (3 \Au^\dag \Au + \An^\dag \An)
\\ &-4\An \Yn^\dag \Tr (3 \Au^\dag \Yu + \An^\dag \Yn)
  -4\Yn \An^\dag \Tr (3 \Yu^\dag \Au + \Yn^\dag \An)
\\& + \Bigl [ 6 g_2^2 + \frac{6}{5} g_1^2 \Bigr ] \Bigl \lbrace
  (\mns+ 2 \mHus) \Yn\Yn^\dag + 2 \Yn \mLs\Yn^\dag
  + \Yn\Yn^\dag \mns + 2 \An \An^\dag \Bigr \rbrace
\\& + 12 g_2^2 \Bigl \lbrace 2 |M_2|^2 \Yn \Yn^\dag -
  M_2^* \An \Yn^\dag - M_2 \Yn \An^\dag \Bigr \rbrace
\\ &+ \frac{12}{5} g_1^2 \Bigl \lbrace 2 |M_1|^2 \Yn \Yn^\dag -
  M_1^* \An \Yn^\dag - M_1 \Yn \An^\dag \Bigr \rbrace\;,
\end{align*}
where we have defined
\begin{align*}
\trym = \:&\mHus - \mHds + \Tr [\mQs - \mLs - 2 \mus+ \mds  + \mes]\;,
\eqnum\displaybreak[2]\\
\sss = \:
&\Tr \Bigl [ \bad
  &-(3 \mHus + \mQs) \Yu^\dag \Yu + 4 \Yu^\dag \mus \Yu
    + (3\mHds - \mQs) \Yd^\dag \Yd - 2 \Yd^\dag \mds \Yd
  \\&+ (\mHds + \mLs) \Ye^\dag \Ye - 2 \Ye^\dag \mes \Ye
    +(-\mHus + \mLs) \Yn^\dag \Yn \Bigr ]
\ead \\&+ \left [ \frac{3}{2} g_2^2 + \frac{3}{10}g_1^2 \right ]
    \left\lbrace \mHus - \mHds - \Tr (\mLs) \right \rbrace
    + \left [ \frac{8}{3} g_3^2 + \frac{3}{2} g_2^2 
    +\frac{1}{30} g_1^2 \right ] \Tr (\mQs )
\\&-\left [ \frac{16}{3} g_3^2 + \frac{16}{15} g_1^2 \right ] \Tr (\mus )
    +\left [ \frac{8}{3} g_3^2 + \frac{2}{15} g_1^2 \right ] \Tr (\mds )
    + \frac{6}{5} g_1^2 \Tr (\mes)\;,
\displaybreak[2]\\
\sigma_1 =\:& \frac{1}{5} g_1^2 \Bigl \lbrace  3 ( \mHus + \mHds ) 
  + \Tr [\mQs + 3 \mLs + 8 \mus + 2 \mds + 6 \mes ]\Bigr \rbrace\;,
\displaybreak[2]\\
\sigma_2 =\:& g_2^2 \left \lbrace
  \mHus + \mHds  + \Tr [3 \mQs +  \mLs ]\right \rbrace\;,
\displaybreak[2]\\
\sigma_3 =\:& g_3^2 \Tr [2 \mQs +  \mus + \mds ]\;.
\end{align*}

Finally, the $\beta$ functions for the trilinear and bilinear
soft terms are:
\begin{align*}
\beto_{\Ad} =
\Ad \biggl\lbrace
&\Tr (3 \Yd \Yd^\dag +\Ye \Ye^\dag) + 5 \Yd^\dag \Yd + \Yu^\dag \Yu
  - \frac{16}{ 3} g_3^2 - 3 g_2^2 - \frac{7}{ 15} g_1^2
  \biggr \rbrace+
\eqnum\\ \Yd \biggl\lbrace
&\Tr(6\Ad \Yd^\dag + 2 \Ae \Ye^\dag) + 4 \Yd^\dag \Ad + 2 \Yu^\dag \Au
\\&+ \frac{32}{ 3} g_3^2 M_3 + 6g_2^2 M_2 + \frac{14}{ 15} g_1^2 M_1
  \biggr\rbrace\;,
\displaybreak[2]\\
\bett_{\Ad} = 
\Ad \biggl\lbrace
&- \Tr (9\Yd \Yd^\dag \Yd \Yd^\dag + 3\Yu \Yd^\dag \Yd \Yu^\dag
  + 3\Ye \Ye^\dag \Ye \Ye^\dag +\Yn \Ye^\dag \Ye \Yn^\dag)
\eqnum\\& - \Yu^\dag \Yu \Tr (3\Yu \Yu^\dag +\Yn \Yn^\dag)
  - 5 \Yd^\dag \Yd \Tr (3 \Yd \Yd^\dag+ \Ye \Ye^\dag)
\\&- 6 \Yd^\dag \Yd \Yd^\dag \Yd- 2 \Yu^\dag \Yu \Yu^\dag \Yu
    - 4 \Yu^\dag \Yu \Yd^\dag \Yd
\\&+ \Bigl [ 16 g_3^2 -\frac{2}{ 5} g_1^2 \Bigr] \Tr(\Yd \Yd^\dag)
    + \frac{6}{ 5} g_1^2 \Tr(\Ye \Ye^\dag)
    + \frac{4}{ 5} g_1^2 \Yu^\dag \Yu
    + \Bigl [12 g^2_2 + \frac{6}{ 5} g_1^2 \Bigr] \Yd^\dag \Yd
\\&-\frac{16}{ 9} g_3^4 + 8 g_3^2 g_2^2 + \frac{8}{ 9} g_3^2 g_1^2
    + \frac{15}{ 2} g_2^4 + g_2^2 g_1^2 + \frac{287}{ 90} g_1^4
    \biggr\rbrace +
\displaybreak[2]\\ \Yd \biggl\lbrace
& - 2\Tr ( \bad 
    &18\Ad \Yd^\dag \Yd \Yd^\dag  + 3\Au \Yd^\dag \Yd \Yu^\dag
      + 3\Ad \Yu^\dag \Yu \Yd^\dag
    \\&+ 6\Ae \Ye^\dag \Ye \Ye^\dag + \An \Ye^\dag \Ye \Yn^\dag
      + \Ae \Yn^\dag \Yn \Ye^\dag) \ead
\\&- \Yu^\dag \Yu \Tr (6\Au \Yu^\dag +2\An \Yn^\dag )
    - 6\Yd^\dag \Yd \Tr (3 \Ad \Yd^\dag + \Ae \Ye^\dag)
\\&- \Yu^\dag \Au \Tr (6\Yu \Yu^\dag +2\Yn \Yn^\dag)
    - 4 \Yd^\dag \Ad \Tr(3 \Yd \Yd^\dag + \Ye \Ye^\dag)
\\&- 6 \Yd^\dag \Yd \Yd^\dag \Ad - 8 \Yd^\dag \Ad \Yd^\dag \Yd
    - 4 \Yu^\dag \Au \Yu^\dag \Yu 
\\& - 4 \Yu^\dag \Yu \Yu^\dag \Au - 4 \Yu^\dag \Au \Yd^\dag \Yd 
    - 2 \Yu^\dag \Yu \Yd^\dag \Ad 
\\&+ \Bigl [ 32 g_3^2 - \frac{4}{ 5} g_1^2 \Bigr] \Tr(\Ad \Yd^\dag)
    + \frac{12}{ 5} g_1^2 \Tr(\Ae \Ye^\dag)
    + \frac{8}{ 5} g_1^2 \Yu^\dag \Au
\\&+ \Bigl [6 g_2^2 + \frac{6}{5} g_1^2 \Bigr] \Yd^\dag \Ad
    - \Bigl [ 32 g_3^2 M_3 - \frac{4}{5} g_1^2 M_1\Bigr]\Tr(\Yd \Yd^\dag)
    - \frac{12}{5} g_1^2 M_1 \Tr(\Ye \Ye^\dag)
\\&- \Bigl [12 g_2^2 M_2 + \frac{8}{5} g_1^2 M_1\Bigr]\Yd^\dag \Yd
    - \frac{8}{5} g_1^2 M_1 \Yu^\dag \Yu
\\&+ \frac{64}{9} g_3^4 M_3 - 16 g_3^2 g_2^2 (M_3 + M_2)
    - \frac{16}{9} g_3^2 g_1^2 (M_3 + M_1)
\\&- {30} g_2^4 M_2 - 2 g_2^2 g_1^2 (M_2 + M_1) - \frac{574}{45} g_1^4 M_1
    \biggr\rbrace\;,
\end{align*}
\begin{align*}
\beto_{\Au} =
\Au \biggl\lbrace 
&\Tr (3\Yu \Yu^\dag +\Yn \Yn^\dag) + 5 \Yu^\dag \Yu + \Yd^\dag \Yd
  - \frac{16}{3} g_3^2 - 3 g_2^2 - \frac{13}{15} g_1^2
  \biggr \rbrace+
\eqnum\\ \Yu \biggl\lbrace
&\Tr(6\Au \Yu^\dag +2\An \Yn^\dag) + 4 \Yu^\dag \Au + 2 \Yd^\dag \Ad
\\&+ \frac{32}{3} g_3^2 M_3 + 6g_2^2 M_2 + \frac{26}{15} g_1^2 M_1
  \biggr \rbrace\;,
\displaybreak[2]\\
\bett_{\Au} = 
\Au \biggl\lbrace
&-\Tr (9\Yu \Yu^\dag \Yu \Yu^\dag +  3\Yu \Yd^\dag \Yd \Yu^\dag
  +3\Yn \Yn^\dag \Yn \Yn^\dag +  \Yn \Ye^\dag \Ye \Yn^\dag)
\eqnum\\&- \Yd^\dag \Yd \Tr (3 \Yd \Yd^\dag+ \Ye \Ye^\dag)
  - 5 \Yu^\dag \Yu \Tr (3\Yu \Yu^\dag +\Yn \Yn^\dag)
\\&- 6 \Yu^\dag \Yu \Yu^\dag \Yu - 2 \Yd^\dag \Yd \Yd^\dag \Yd
     - 4 \Yd^\dag \Yd \Yu^\dag \Yu
\\&+ \Bigl [ 16 g_3^2 + \frac{4}{ 5} g_1^2 \Bigr] \Tr(\Yu \Yu^\dag)
    + 12 g_2^2 \Yu^\dag \Yu + \frac{2}{ 5} g_1^2 \Yd^\dag \Yd
\\&-\frac{16}{ 9} g_3^4 + 8 g_3^2 g_2^2 + \frac{136}{ 45} g_3^2 g_1^2
    + \frac{15}{ 2} g_2^4 + g_2^2 g_1^2 + \frac{2743}{ 450} g_1^4
    \biggr\rbrace +
\displaybreak[2]\\ \Yu \biggl\lbrace
& - 2\Tr ( \bad 
  & 18\Au \Yu^\dag \Yu \Yu^\dag + 3\Au \Yd^\dag \Yd \Yu^\dag
      + 3\Ad \Yu^\dag \Yu \Yd^\dag
  \\&+6\An \Yn^\dag \Yn \Yn^\dag      +\An \Ye^\dag \Ye \Yn^\dag
      +\Ae \Yn^\dag \Yn \Ye^\dag)
\ead\\&- 6\Yu^\dag \Yu \Tr (3\Au \Yu^\dag +\An \Yn^\dag)
  - \Yd^\dag \Yd \Tr(6 \Ad \Yd^\dag + 2 \Ae \Ye^\dag)
\\&- 4 \Yu^\dag \Au \Tr (3\Yu \Yu^\dag+\Yn \Yn^\dag)
  - \Yd^\dag \Ad \Tr (6 \Yd \Yd^\dag + 2 \Ye \Ye^\dag)
\\&- 6 \Yu^\dag \Yu \Yu^\dag \Au- 8 \Yu^\dag \Au \Yu^\dag \Yu
    - 4 \Yd^\dag \Yd \Yd^\dag \Ad
\\&- 4 \Yd^\dag \Ad \Yd^\dag \Yd - 2 \Yd^\dag \Yd \Yu^\dag \Au
    - 4 \Yd^\dag \Ad \Yu^\dag \Yu
\\&+ \Bigl [ 32 g_3^2 + \frac{8}{  5} g_1^2 \Bigr] \Tr(\Au \Yu^\dag)
    + \Bigl [6 g_2^2 + \frac{6 }{ 5} g_1^2\Bigr] \Yu^\dag \Au
    + \frac{4}{ 5} g_1^2 \Yd^\dag \Ad 
\\&- \Bigl [ 32 g_3^2 M_3 + \frac{8}{  5} g_1^2 M_1\Bigr] \Tr(\Yu \Yu^\dag)
    - \Bigl [12 g_2^2 M_2+ \frac{4}{ 5} g_1^2 M_1\Bigr] \Yu^\dag \Yu 
\\&- \frac{4}{5} g_1^2 M_1 \Yd^\dag \Yd
    + \frac{64}{9} g_3^4 M_3 - 16 g_3^2 g_2^2 (M_3 + M_2)
    - \frac{272}{45} g_3^2 g_1^2 (M_3 + M_1)
\\&- 30 g_2^4 M_2 - 2 g_2^2 g_1^2 (M_2 + M_1) - \frac{5486}{ 225} g_1^4 M_1
    \biggr\rbrace\;,
\end{align*}
\begin{align*}
\beto_{\Ae} = 
\Ae \biggl\lbrace 
&\Tr (3 \Yd \Yd^\dag + \Ye \Ye^\dag)
  + 5 \Ye^\dag \Ye +\Yn^\dag \Yn
  - 3 g_2^2 - \frac{9}{5} g_1^2
  \biggr \rbrace + 
\eqnum\\ \Ye \biggl\lbrace
&\Tr(6 \Ad \Yd^\dag + 2\Ae \Ye^\dag)
  + 4 \Ye^\dag \Ae+ 2 \Yn^\dag \An + 6g_2^2 M_2 + \frac{18}{5} g_1^2 M_1
  \biggr \rbrace\;,
\displaybreak[2]\\
\bett_{\Ae} =
\Ae \biggl\lbrace
&- \Tr (9\Yd \Yd^\dag \Yd \Yd^\dag  + 3\Yu \Yd^\dag \Yd \Yu^\dag
     + 3\Ye \Ye^\dag \Ye \Ye^\dag +\Yn \Ye^\dag \Ye \Yn^\dag)
\eqnum\\&- \Yn^\dag \Yn \Tr (3\Yu \Yu^\dag +\Yn \Yn^\dag)
    - 5 \Ye^\dag \Ye \Tr (3 \Yd \Yd^\dag + \Ye \Ye^\dag)
\\&- 6 \Ye^\dag \Ye \Ye^\dag \Ye - 2 \Yn^\dag \Yn \Yn^\dag \Yn
    - 4 \Yn^\dag \Yn \Ye^\dag \Ye
\\&+ \Bigl [ 16 g_3^2 - \frac{2}{5} g_1^2 \Bigr] \Tr(\Yd \Yd^\dag)
    + \frac{6}{5} g_1^2 \Tr(\Ye \Ye^\dag)
    + \Bigl [12 g_2^2 - \frac{6}{5} g_1^2 \Bigr] \Ye^\dag \Ye
\\&+ \frac{15}{2} g_2^4 + \frac{9}{5}g_2^2 g_1^2 + \frac{27}{2} g_1^4
    \biggr\rbrace +
\displaybreak[2]\\ \Ye \biggl\lbrace
& - 2\Tr ( \bad
    &18\Ad \Yd^\dag \Yd \Yd^\dag + 3\Au \Yd^\dag \Yd \Yu^\dag
      + 3\Ad \Yu^\dag \Yu \Yd^\dag\\
    & + 6\Ae \Ye^\dag \Ye \Ye^\dag+ \An \Ye^\dag \Ye \Yn^\dag
      + \Ae \Yn^\dag \Yn \Ye^\dag)
\ead \\&- \Yn^\dag \Yn \Tr (6\Au \Yu^\dag +2\An \Yn^\dag )
    - 6\Ye^\dag \Ye \Tr (3 \Ad \Yd^\dag+ \Ae \Ye^\dag)
\\&- \Yn^\dag \An \Tr (6\Yu \Yu^\dag+ 2\Yn \Yn^\dag)
    - 4 \Yd^\dag \Ad \Tr(3 \Yd \Yd^\dag + \Ye \Ye^\dag)
\\&- 6 \Ye^\dag \Ye \Ye^\dag \Ae  - 8 \Ye^\dag \Ae \Ye^\dag \Ye
    - 4 \Yn^\dag \An \Yn^\dag \Yn
\\&- 4 \Yn^\dag \Yn \Yn^\dag \An  - 4 \Yn^\dag \An \Ye^\dag \Ye
    - 2 \Yn^\dag \Yn \Ye^\dag \Ae
\\&+\Bigl [ 32 g_3^2 - \frac{4}{5} g_1^2 \Bigr] \Tr(\Ad \Yd^\dag)
    + \frac{12}{5} g_1^2 \Tr(\Ae \Ye^\dag)
    + \Bigl [6 g_2^2 + \frac{6}{5} g_1^2 \Bigr] \Ye^\dag \Ae
\\&- \Bigl [ 32 g_3^2 M_3 - \frac{4}{5} g_1^2 M_1\Bigr] \Tr(\Yd \Yd^\dag)
    - \frac{12}{5} g_1^2 M_1 \Tr(\Ye \Ye^\dag) -12 g_2^2 M_2 \Ye^\dag \Ye
\\&- {30} g_2^4 M_2 -\frac{18}{5}g_2^2 g_1^2 (M_1 + M_2) - 54 g_1^4 M_1
    \biggr\rbrace\;,
\end{align*}
\begin{align*}
\beto_{\An}  = 
\An \biggl\lbrace
&\Tr (3\Yu \Yu^\dag +\Yn \Yn^\dag) + 5 \Yn^\dag \Yn
  + \Ye^\dag \Ye - 3 g_2^2 - \frac{3}{5} g_1^2
  \biggr \rbrace +
\eqnum\\ \Yn \biggl\lbrace
&\Tr(6\Au \Yu^\dag + 2\An \Yn^\dag) + 4 \Yn^\dag \An
  + 2 \Ye^\dag \Ae + 6 g_2^2 M_2 + \frac{6}{5} g_1^2 M_1
  \biggr \rbrace\;,
\displaybreak[2]\\
\bett_{\An}  = 
\An \biggl\lbrace
&-\Tr (9\Yu \Yu^\dag \Yu \Yu^\dag +  3\Yu \Yd^\dag \Yd \Yu^\dag 
    +3\Yn \Yn^\dag \Yn \Yn^\dag + \Yn \Ye^\dag \Ye \Yn^\dag)
\\&- \Ye^\dag \Ye \Tr (3 \Yd \Yd^\dag+ \Ye \Ye^\dag)
    - 5 \Yn^\dag \Yn \Tr (3\Yu \Yu^\dag +\Yn \Yn^\dag)
\eqnum\\&- 6 \Yn^\dag \Yn \Yn^\dag \Yn  - 2 \Ye^\dag \Ye \Ye^\dag \Ye
    - 4 \Ye^\dag \Ye \Yn^\dag \Yn
\\&+ \Bigl [ 16 g_3^2 + \frac{4}{5} g_1^2 \Bigr] \Tr(\Yu \Yu^\dag)
    + \Bigl [ 12 g_2^2 + \frac{12}{5} g_1^2 \Bigr] \Yn^\dag \Yn
    + \frac{6}{5} g_1^2 \Ye^\dag \Ye
\\&+ \frac{15}{2} g_2^4 +\frac{9}{5} g_2^2 g_1^2 + \frac{207}{50} g_1^4
    \biggr\rbrace +
\displaybreak[2]\\ \Yn \biggl\lbrace
& - 2\Tr (\bad
    & 18\Au \Yu^\dag \Yu \Yu^\dag + 3\Au \Yd^\dag \Yd \Yu^\dag
      + 3\Ad \Yu^\dag \Yu \Yd^\dag
    \\&+6\An \Yn^\dag \Yn \Yn^\dag + \An \Ye^\dag \Ye \Yn^\dag
      + \Ae \Yn^\dag \Yn \Ye^\dag) \ead
\\&- 6 \Yn^\dag \Yn \Tr (3\Au \Yu^\dag +\An \Yn^\dag)
    - \Ye^\dag \Ye \Tr (6 \Ad \Yd^\dag + 2 \Ae \Ye^\dag)
\\&- 4 \Yn^\dag \An \Tr (3\Yu \Yu^\dag +\Yn \Yn^\dag)
    - \Ye^\dag \Ae \Tr (6 \Yd \Yd^\dag + 2 \Ye \Ye^\dag)
\\&- 6 \Yn^\dag \Yn \Yn^\dag \An - 8 \Yn^\dag \An \Yn^\dag \Yn
    - 4 \Ye^\dag \Ye \Ye^\dag \Ae - 4 \Ye^\dag \Ae \Ye^\dag \Ye
\\&- 2 \Ye^\dag \Ye \Yn^\dag \An - 4 \Ye^\dag \Ae \Yn^\dag \Yn
    + \Bigl [ 32 g_3^2 + \frac{8}{5} g_1^2 \Bigr] \Tr(\Au \Yu^\dag) 
\\&+ \Bigl [6 g_2^2 + \frac{6}{5} g_1^2\Bigr] \Yn^\dag \An
    + \frac{12}{5} g_1^2 \Ye^\dag \Ae
    - \Bigl [ 32 g_3^2 M_3 + \frac{8}{5} g_1^2 M_1\Bigr] \Tr(\Yu \Yu^\dag)
\\&- \Bigl [12 g_2^2 M_2+ \frac{12}{5} g_1^2 M_1\Bigr] \Yn^\dag \Yn
    - \frac{12}{5} g_1^2 M_1 \Ye^\dag \Ye 
\\&- 30 g_2^4 M_2 - \frac{18}{5} g_2^2 g_1^2 (M_2 + M_1) - \frac{414}{25} g_1^4 M_1
    \biggr\rbrace\;,
\end{align*}
\begin{align*}
\beto_B = 
B \biggl \lbrace
&\Tr (3 \Yu \Yu^\dag +3  \Yd \Yd^\dag + \Yn \Yn^\dag + \Ye \Ye^\dag )
  - 3 g_2^2 - \frac{3}{5} g_1^2 
  \biggr \rbrace +
\eqnum\\\mu \biggl \lbrace
&\Tr ( 6 \Au \Yu^\dag + 6 \Ad \Yd^\dag + 2\An \Yn^\dag + 2\Ae \Ye^\dag)
  + 6 g_2^2 M_2 +\frac{6}{5} g_1^2 M_1
  \biggr \rbrace\;,
\displaybreak[2]\\
\bett_B =
B \biggl\lbrace
&- \Tr ( \eqnum \bad
    &9\Yu \Yu^\dag \Yu \Yu^\dag +  9\Yd \Yd^\dag \Yd \Yd^\dag
      + 6\Yu \Yd^\dag \Yd \Yu^\dag
    \\&+3\Yn \Yn^\dag \Yn \Yn^\dag  +  3\Ye \Ye^\dag \Ye \Ye^\dag
  +  2 \Yn \Ye^\dag \Ye \Yn^\dag)
\ead \\&+ \Bigl [16 g_3^2 + \frac{4}{5} g_1^2 \Bigr] \Tr(\Yu \Yu^\dag )
    + \Bigl [16 g_3^2 - \frac{2}{5} g_1^2 \Bigr] \Tr(\Yd \Yd^\dag )
    + \frac{6}{5} g_1^2 \Tr(\Ye \Ye^\dag ) 
\\&+ \frac{15}{2} g_2^4 + \frac{9}{5} g_1^2 g_2^2 + \frac{207}{50} g_1^4
    \biggr\rbrace +
\displaybreak[2]\\ \mu \biggl\lbrace
&- 4 \Tr (\bad
    &9 \Au \Yu^\dag \Yu \Yu^\dag + 9 \Ad \Yd^\dag \Yd \Yd^\dag
      + 3 \Au \Yd^\dag \Yd \Yu^\dag + 3 \Ad \Yu^\dag \Yu \Yd^\dag 
    \\&+3\An \Yn^\dag \Yn \Yn^\dag + 3 \Ae \Ye^\dag \Ye \Ye^\dag
      +  \An \Ye^\dag \Ye \Yn^\dag +
      \Ae \Yn^\dag \Yn \Ye^\dag) 
\ead \\&+ \Bigl [32 g_3^2 + \frac{8}{5} g_1^2 \Bigr] \Tr(\Au \Yu^\dag )
    + \Bigl [32 g_3^2 - \frac{4}{5} g_1^2 \Bigr] \Tr(\Ad \Yd^\dag )
    + \frac{12}{5} g_1^2 \Tr(\Ae \Ye^\dag ) 
\\&- \Bigl [32 g_3^2 M_3 + \frac{8}{5} g_1^2 M_1 \Bigr] \Tr(\Yu \Yu^\dag )
    - \Bigl [32 g_3^2 M_3 - \frac{4}{5} g_1^2 M_1 \Bigr] \Tr(\Yd \Yd^\dag ) 
\\&- \frac{12}{5} g_1^2 M_1 \Tr(\Ye \Ye^\dag ) - {30} g_2^4 M_2 - 
    \frac{18}{5} g_1^2 g_2^2 (M_1 + M_2) - \frac{414}{25}g_1^4 M_1
    \biggr\rbrace\;,
\end{align*}
\begin{align*}
\beto_\BM = \BM \biggl [& 2 \Yn^\con \Yn^\tra \biggr ] +\sM \biggl [4 \Yn^\con \An^\tra \biggr ]+ \biggl [2\Yn \Yn^\dag \biggr ] \BM + \biggl [4\An \Yn^\dag \biggr ] \sM
\;,
\eqnum\displaybreak[2]\\
\bett_\BM  = \BM \eqnum  \biggl [ 
&-2 \Yn^\con \Ye^\tra \Ye^\con \Yn^\tra -2 \Yn^\con \Yn^\tra \Yn^\con \Yn^\tra
  -2 \Yn^\con \Yn^\tra \Tr (3 \Yu \Yu^\dag + \Yn \Yn^\dag) 
\\& +\frac{6}{5} g_1^2 \Yn^\con \Yn^\tra + 6 g_2^2 \Yn^\con \Yn^\tra \biggr ]+
\\ \sM \biggl [
&-4 \Yn^\con \Ae^\tra \Ye^\con \Yn^\tra -4 \Yn^\con \An^\tra \Yn^\con \Yn^\tra
  -4 \Yn^\con \Yn^\tra \Tr (3 \Au \Yu^\dag + \An \Yn^\dag) 
\\&-4 \Yn^\con \Ye^\tra \Ye^\con \An^\tra -4 \Yn^\con \Yn^\tra \Yn^\con \An^\tra
  -4 \Yn^\con \An^\tra \Tr (3 \Yu \Yu^\dag + \Yn \Yn^\dag) 
\\& +\frac{12}{5} g_1^2 \Yn^\con \An^\tra + 12 g_2^2 \Yn^\con \An^\tra
  -\frac{12}{5} g_1^2 M_1 \Yn^\con \Yn^\tra - 12 g_2^2 M_2 \Yn^\con \Yn^\tra \biggr]+
\\ \biggl [
&-2 \Yn \Ye^\dag \Ye \Yn^\dag -2 \Yn \Yn^\dag \Yn \Yn^\dag
   -2 \Yn \Yn^\dag \Tr (3 \Yu \Yu^\dag + \Yn \Yn^\dag)
\\&+\frac{6}{5} g_1^2 \Yn \Yn^\dag + 6 g_2^2 \Yn \Yn^\dag
\biggr ] \BM+
\\ \biggl [
&-4 \Yn \Ye^\dag \Ae \Yn^\dag -4 \Yn \Yn^\dag \An \Yn^\dag
   -4 \Yn \Yn^\dag \Tr (3 \Au \Yu^\dag + \An \Yn^\dag)
\\&-4 \An \Ye^\dag \Ye \Yn^\dag -4 \An \Yn^\dag \Yn \Yn^\dag
   -4 \An \Yn^\dag \Tr (3 \Yu \Yu^\dag + \Yn \Yn^\dag)
\\&+\frac{12}{5} g_1^2 \An \Yn^\dag + 12 g_2^2 \An \Yn^\dag
   -\frac{12}{5} g_1^2 M_1 \Yn \Yn^\dag -12 g_2^2 M_2 \Yn \Yn^\dag 
\biggr ] \sM\;.
\end{align*}

\end{document}